\documentclass[aps,prb,twocolumn,superscriptaddress,longbibliography,nofootinbib]{revtex4-2}
\usepackage{amsmath,amssymb,graphicx,bm,xcolor}
\usepackage[colorlinks=true,linkcolor=blue,citecolor=blue,urlcolor=blue]{hyperref}
\newcommand{\morb}{\mathcal M}
\newcommand{\cchi}{\mathcal X}
\newcommand{\eeff}{\hbar\tilde\omega_{\mathrm{eff}}}
\newcommand{\Atil}{\tilde A}
\newcommand{\Aocc}{A_{\mathrm{occ}}}
\newcommand{\AM}{A_{\mathcal M}}
\newcommand{\FT}{\mathcal F_T}

\begin{document}

\title{Orbital magnetic susceptibility and de Haas--van Alphen effect of a flat band from quantum geometry}

\author{Ethan Huecker}
\affiliation{Department of Physics, University of Florida, Gainesville, Florida 32611, USA}

\author{Mengxing Ye}
\affiliation{Department of Physics and Astronomy, University of Utah, Salt Lake City, Utah 84112, USA}

\author{Yuxuan Wang}
\affiliation{Department of Physics, University of Florida, Gainesville, Florida 32611, USA}
\affiliation{Anthropic STEM Fellows Program, San Francisco, CA, USA}

\date{\today}

\begin{abstract}
A band that is exactly flat at zero magnetic field has zero velocity, a divergent effective mass, and no Fermi
surface, so the standard theory of orbital magnetism does not
apply. We show that the magnetic field generates an effective dispersion.
In the expansion organized by the Moyal algebra of the
noncommutative kinetic momenta, the band energy acquires a
first-order shift $B\,\morb(\mathbf k)$, set by the orbital moment
of the Bloch state, and a second-order shift
$\tfrac12\cchi(\mathbf k)B^{2}$, whose gauge-invariant coefficient
we obtain in closed form. These two functions, together with the Berry
curvature $\Omega$, fix the entire orbital response. The
susceptibility is purely geometric, and for a generic isolated flat
band it reproduces the established gauge-invariant response formula. With the chemical potential pinned to the flat band, the
$\morb=0$ contour becomes an emergent Fermi surface, and
bosonizing excitations in its vicinity yields a modified
Lifshitz--Kosevich formula with level spacing quadratic in $B$. The phase shift of the oscillations is given jointly by the Berry
phase around the emergent Fermi surface and $\cchi$. When the band is narrow rather than exactly flat, the oscillation
frequency crosses over, with a parameter-free profile, from the
zero-field Fermi-surface area to the area of the $\morb=0$
contour. We verify our analytical findings
numerically by exactly diagonalizing a flattened Chern lattice model.
\end{abstract}

\maketitle

\section{Introduction}
\label{sec:intro}

The orbital magnetic response of a metal is organized by its Fermi
surface. The Landau--Peierls susceptibility is built from band
curvature at the Fermi level \cite{Peierls1933}, and the de Haas--van
Alphen (dHvA) effect measures extremal Fermi-surface areas through
the Onsager relation \cite{Onsager1952}, with amplitudes given by
Lifshitz--Kosevich (LK) theory \cite{LK1956,Shoenberg1984}. An
exactly flat band has none of these ingredients. The effective mass
diverges, the group velocity vanishes, and at any partial filling
there is no Fermi contour to quantize. Such bands are now routinely
engineered, in twisted bilayer graphene near the magic angle
\cite{BM2011,CaoMott2018,CaoSC2018}, in line-graph and kagome
lattices \cite{RhimKimYang2020}, and in further moir\'e materials,
and in them only the wavefunctions depend on momentum. Their quantum
geometry, described by the Berry curvature and the quantum
metric, controls the superfluid weight
\cite{PeottaTorma2015,TormaPeottaBernevig2022} and the structure of
the Landau levels \cite{RhimKimYang2020,RhimYang2021review}.

For an isolated flat band, Ref.~\cite{HwangRhimYang2021} showed
that the Landau levels are those of the effective dispersion
$B\,\morb(\mathbf k)$ set by the orbital moment of the Bloch
states, with level spacing quadratic in $B$. For a special
class of exactly flat Chern bands with ideal quantum geometry,
Ref.~\cite{Liu2026} obtained closed-form Landau levels at all field
strengths, and from them the quantum-oscillation spectrum. For a
generic isolated flat band, what these levels imply for the dHvA
effect, its frequency, amplitudes, and phase shift, remains an
open question, and first order in $B$ is not enough to answer it. The phase of the oscillations is an orbit area divided
by $B$. A second-order term of the dispersion shifts the area at
first order and so enters the phase at order one. There is also no
guarantee that the semiclassical rules for a metal, Onsager
quantization with its Berry-phase and Roth corrections
\cite{MikitikSharlai1999,Roth1966}, carry over to a dispersion the
field itself generates. In parallel, the orbital magnetism of Bloch electrons, a classic
subject \cite{Fukuyama1971}, has come under intense investigation
from the perspectives of quantum geometry
\cite{Xiao2010,Raoux2015,Piechon2016,GaoNiu2017} and of electron
interactions \cite{KWV2025,Ye2026,MayrhoferChubukov2023}. These
works compute the magnetization and the susceptibility, the first
two derivatives of the free energy in the field. The dHvA
oscillations are invisible to this expansion. At fixed temperature
they vanish faster than any power of $B$ as $B\to0$, contributing
to no order in the field. A unified theory that produces the smooth
derivatives and the oscillations from the same starting point has
been missing. Given only the zero-field wavefunctions of an
isolated flat band, what is the full orbital response, smooth and
oscillatory?

In this paper we derive this theory to second order in the field. At first order, each momentum state of the flat band acquires the energy
$B\,\morb(\mathbf k)$ of its own orbital moment. At second order a
coefficient $\cchi(\mathbf k)$ enters, for which we derive a closed
form from a systematic Moyal expansion of the magnetic Bloch
problem. The field-generated band
\begin{equation}
\varepsilon(\mathbf k;B)=\varepsilon_0
+ B\,\morb(\mathbf k) + \frac{1}{2}\,B^{2}\cchi(\mathbf k)
\label{eq:introband}
\end{equation}
then feeds the standard thermodynamic machinery, with one geometric
addition: the phase-space measure acquires the Berry factor
$1+B\Omega$ \cite{Xiao2010}. The smooth response and the oscillations both
follow from this band. All results below are expressed in these
three functions of $\mathbf k$, evaluated on the zero-field
wavefunctions.

The susceptibility is a purely geometric functional of $\morb$,
$\cchi$, and $\Omega$, and for a generic isolated flat band we
prove that it agrees with the exact resolvent formula of
Ref.~\cite{Raoux2015}. With the chemical potential pinned to the
flat band, the $\morb=0$ contour
becomes an emergent Fermi surface, and bosonizing the excitations in
its vicinity along the lines of Ref.~\cite{YeWang2025} produces a
modified Lifshitz--Kosevich series whose quantization phase carries
an $O(1)$ contribution from $\cchi$ alongside the Berry phase. On the hole pocket of the lattice model of
Sec.~\ref{sec:numerics}, we find that the contribution from $\cchi$ is larger than the Berry phase. The form of the series is obtained
without invoking the semiclassical quantization rule of cyclotron
orbits. When the band is narrow rather than exactly flat, Landau quantization proceeds on contours of the composite
dispersion, built-in plus field-generated, and the dHvA frequency
crosses over from the zero-field Fermi-surface
area to the area of the $\morb=0$ contour, with a parameter-free
profile and a $1/B$ asymptotic behavior.

The expansion is controlled by the single parameter $x=B\xi^{2}$,
with $\xi$ the geometric length of the band. Each order of
the Moyal expansion carries one factor of $x$, and the closed forms
for $\morb$ and $\cchi$ confirm the counting
(Sec.~\ref{sec:control}). The derivation applies to an isolated,
nondegenerate flat band whose moment $\morb(\mathbf k)$ is smooth
with a single closed $\morb=0$ contour, with no condition on its
quantum geometry. These assumptions are collected
in Sec.~\ref{sec:LK} where each is used. The scope is
complementary to that of Ref.~\cite{Liu2026}, which is exact in $B$
but restricted to bands with ideal quantum geometry, while our
expansion works to $O(B^{2})$. Where the two approaches overlap, they
agree: our expansion reproduces the weak-field sector of that
solvable model.

The oscillatory part of the analysis uses coadjoint-orbit
bosonization, an exact bosonic description of a Fermi surface in
terms of its shape fluctuations \cite{DelacretazDuMehtaSon2022}. It
was applied to the de Haas--van Alphen effect of a Fermi gas and a
Fermi liquid in Ref.~\cite{Wang2025}, and extended in
Ref.~\cite{YeWang2025} to multiband systems, where the Berry
curvature enters through the phase-space measure. We extend the
construction of Ref.~\cite{YeWang2025}, which keeps the dispersion to
first order in the field, to the full second-order dispersion. For a
dispersive band the second-order term shifts the quantization phase
only at $O(B)$. For a flat band, whose level spacing is quadratic in
$B$, it enters at order one.

The paper is organized as follows. Section~\ref{sec:dispersion}
constructs the field-generated dispersion and establishes its gauge
structure. Section~\ref{sec:chi} derives the susceptibility. Its
consistency with the established gauge-invariant response theory is
proved for a generic flat band in Appendix~\ref{app:consistency}.
Section~\ref{sec:LK} derives the LK formula from bosonization.
Section~\ref{sec:numerics} checks the analytic results against exact
diagonalization of a flattened Chern lattice model.
Section~\ref{sec:crossover} extends the theory to narrow bands and
derives the crossover of the dHvA frequency.
Section~\ref{sec:conclusions} concludes with the experimental
outlook.

\emph{Notation.} $\morb(\mathbf k)$ is the orbital moment of the Bloch
state and $\cchi(\mathbf k)$ the second-order dispersion coefficient;
$\Omega(\mathbf k)$ is the Berry curvature; $M$ denotes the
thermodynamic magnetization; $\Xi$ is the grand potential;
$\Delta$ is the \emph{full} gap between the flat band and its nearest
neighbor; $\Atil(\morb')$ is the area enclosed by the moment contour
$\morb(\mathbf k)=\morb'$, $\AM=\Atil(0)$ the area enclosed by the
$\morb=0$ contour $\mathcal C_0$, and $\Aocc$ the area of the occupied region.
Band indices appear as $b\neq v$ in sums over remote bands, where
$v$ labels the flat band and $b$ the remote bands. We set
$\hbar=e=1$. Then $B$ is the magnetic flux per unit cell, and the flux
quantum is $2\pi$.

\section{The field-generated dispersion}
\label{sec:dispersion}

\subsection{Moyal expansion}
\label{sec:moyal}

Take an isolated band that is exactly flat at $B=0$,
$\varepsilon_0(\bm\pi)=\varepsilon_0$, where $\bm\pi$ is the kinetic
momentum: the argument of the zero-field Bloch Hamiltonian, promoted
in a uniform field to the noncommuting pair
$[\pi_x,\pi_y]=+iB$, and reducing to the crystal momentum as
$B\to0$. Switching on a perpendicular field
generates a dispersion whose only possible origin is the
momentum dependence of the wavefunctions. This section computes it
to second order. Once both coefficients are in closed form,
estimating their magnitudes by counting wavefunction derivatives and
comparing successive orders yields the expansion parameter
$x=B\xi^{2}$ (Sec.~\ref{sec:control}).

Because $\pi_x$ and $\pi_y$ do not commute, $\widehat H(\bm\pi)$
is a function of noncommuting arguments, and products of such
functions are operator products. The
standard remedy is to represent each operator by its Weyl symbol, an
ordinary function of $\bm\pi$, with the operator product encoded in
the Moyal product of the symbols. The magnetic Bloch problem for the
periodic part of the wavefunction then becomes a star-eigenvalue
problem \cite{Blount1962,Roth1962},
\begin{equation}
\widehat H \star |v\rangle = |v\rangle \star \varepsilon,
\qquad
\star = \exp\!\Big[+\frac{iB}{2}\,\epsilon_{ij}\,
\overleftarrow{\partial}_i\overrightarrow{\partial}_j\Big],
\label{eq:star}
\end{equation}
where $\widehat H(\bm\pi)$ is the zero-field Bloch Hamiltonian
evaluated at the kinetic momentum, $\partial_i=\partial/\partial\pi_i$
here and throughout this section, and the arrows indicate which
factor each derivative acts on: $\overleftarrow{\partial}_i$
differentiates the symbol on the left of the $\star$,
$\overrightarrow{\partial}_j$ the symbol on the right. The star
product organizes the field corrections order by order in $B$.
We expand state and eigenvalue in powers of $B$,
\begin{equation}
|v\rangle = |v^{0}\rangle + |v^{1}\rangle + \cdots,
\quad
\varepsilon = \varepsilon_0 + \varepsilon^{(1)}
+ \varepsilon^{(2)} + \cdots,
\label{eq:expand}
\end{equation}
where $|v^{n}\rangle$ and $\varepsilon^{(n)}$ collect the $O(B^{n})$ contributions, and projecting order by order gives them as explicit
functionals of the zero-field band.

\subsection{First order: the orbital moment}
\label{sec:first}

Since $\varepsilon_0$ is a constant, all its derivatives vanish and
it star-multiplies as an ordinary number, so, using the
normalization $\langle v|\!\star\!|v\rangle=1$, the projected
dispersion $\varepsilon=\langle v|\!\star\!\widehat H\!\star\!|v\rangle$
splits exactly into
\begin{equation}
\varepsilon = \varepsilon_0\,\langle v|\!\star\!|v\rangle
+ \langle v|\!\star\!(\widehat H-\varepsilon_0)\!\star\!|v\rangle .
\label{eq:epsplit}
\end{equation}
The normalization $\langle v|\!\star\!|v\rangle=1$ fixes the first
term to the constant $\varepsilon_0$. The field-generated
dispersion comes entirely from the gap-weighted second term, at every order, since a dispersion cannot depend on the value of the constant $\varepsilon_0$.

At $O(B)$ the cross terms
$\langle v^{0}|(\widehat H-\varepsilon_0)|v^{1}\rangle$ and its
conjugate vanish, because $(\widehat H-\varepsilon_0)$ annihilates
$|v^{0}\rangle$ whether it acts to the right or to the left. What
remain are the $\langle v^{0}|\cdots|v^{0}\rangle$ terms, where
each Moyal product in the second term contributes one derivative
insertion $\tfrac{iB}{2}\epsilon_{ij}\,\partial_i\!\cdot\partial_j$.
The insertion at the left Moyal product differentiates the product
$(\widehat H-\varepsilon_0)|v^{0}\rangle$ and splits in two by the
product rule, giving three matrix elements in all. A useful
relation, obtained by differentiating the zero-field eigenvalue
equation $\widehat H|v^{0}\rangle=\varepsilon_0|v^{0}\rangle$,
\begin{equation}
(\partial_i\widehat H)|v^{0}\rangle
= -(\widehat H-\varepsilon_0)|\partial_i v^{0}\rangle,
\label{eq:id1}
\end{equation}
together with its Hermitian conjugate, converts the two matrix elements that carry $\partial\widehat H$,
and the dispersion reduces as follows:
\begin{align}
\varepsilon^{(1)}
&=\frac{iB}{2}\,\epsilon_{ij}\Big[
\langle\partial_i v^{0}|(\partial_j\widehat H)|v^{0}\rangle
+\langle\partial_i v^{0}|(\widehat H-\varepsilon_0)|\partial_j v^{0}\rangle
\nonumber\\
&\hspace{5.4em}
+\langle v^{0}|(\partial_i\widehat H)|\partial_j v^{0}\rangle
\Big]
\nonumber\\
&=\frac{iB}{2}\,\epsilon_{ij}\,(-1+1-1)\,
\langle\partial_i v^{0}|(\widehat H-\varepsilon_0)|\partial_j v^{0}\rangle
\nonumber\\
&= B\,\morb(\bm\pi),
\label{eq:firstchain}
\end{align}
where
\begin{equation}
\morb \equiv -\frac{i}{2}\,\epsilon_{ij}\,
\langle\partial_i v^{0}|\,(\widehat H-\varepsilon_0)\,|\partial_j v^{0}\rangle .
\label{eq:morb}
\end{equation}
Equation~\eqref{eq:morb} is the
orbital magnetic moment of the Bloch state
\cite{Blount1962,Roth1962,GaoNiu2017,Xiao2010} in our sign
conventions. It is gauge invariant as written. Under
$|v^{0}\rangle\to e^{i\phi}|v^{0}\rangle$ the extra terms all
carry $(\widehat H-\varepsilon_0)|v^{0}\rangle$ or its conjugate,
both zero.

To leading order, then, the field generates a
dispersive band due to the momentum-dependent orbital moment, with
bandwidth $(\morb^{\max}-\morb^{\min})\,B$ linear in $B$.
Wherever $\morb$ changes sign, the field-generated
band straddles the flat-band energy, and the level set
\begin{equation}
\mathcal C_0 \;=\; \{\bm\pi:\ \morb(\bm\pi)=0\}
\label{eq:zerocontour}
\end{equation}
becomes a Fermi surface for a chemical potential pinned at $\varepsilon_0$. Equation \eqref{eq:zerocontour} is an emergent Fermi contour, present only at finite field, whose enclosed area is determined by the zero-field wavefunctions. A sign change of $\morb$ is not guaranteed. A
band whose moment has a definite sign, as symmetry may enforce, has
no such contour, and the field then shifts the band without creating
a Fermi surface. In this work we focus on the case where the
contour exists.

\subsection{Second order: the closed form of $\cchi$}
\label{sec:second}

At $O(B^{2})$ the advantage of the split \eqref{eq:epsplit}
remains. The raw expansion contains the unknown
second-order state $|v^{2}\rangle$, but it enters the gap-weighted
term only through
$\langle v^{0}|(\widehat H-\varepsilon_0)|v^{2}\rangle$ and its
conjugate, both zero, so $|v^{2}\rangle$ drops out identically and
never needs to be computed. The second term of
Eq.~\eqref{eq:epsplit} is expanded at $O(B^{2})$ by distributing the
derivative insertions of the two star products over the state
expansion \eqref{eq:expand},
\begin{align}
\varepsilon^{(2)} ={}&
\Big[\langle v^{0}|\!\star\!(\widehat H-\varepsilon_0)\!\star\!|v^{0}\rangle\Big]_{O(B^{2})}
\nonumber\\
&+2\,\mathrm{Re}\Big[\langle v^{0}|\!\star\!(\widehat H-\varepsilon_0)\!\star\!|v^{1}\rangle\Big]_{O(B)}
\nonumber\\
&+\langle v^{1}|(\widehat H-\varepsilon_0)|v^{1}\rangle ,
\label{eq:new-secondsplit}
\end{align}
where the subscript selects the stated order of the star insertions
$\tfrac{iB}{2}\epsilon_{ij}\partial_i\!\cdot\partial_j$ and the
$\langle v^{0}|\cdots|v^{2}\rangle$ terms have already been dropped.
The first bracket carries two insertions, both at the same star
product, which contributes with weight $\tfrac12(\tfrac{iB}{2})^{2}$
from the exponential, or one at each, with weight
$(\tfrac{iB}{2})^{2}$. The product rule distributes the derivatives
over $(\widehat H-\varepsilon_0)$ and the states, and the identity
\eqref{eq:id1} removes the terms in which no derivative acts on
$|v^{0}\rangle$. The second bracket carries one insertion.
Its surviving terms reduce under Eq.~\eqref{eq:id1} to a single
$\mathrm{Im}$ matrix element. Carrying out this bookkeeping leaves
\begin{align}
\varepsilon^{(2)} ={}&
-\frac{B^{2}}{8}\,\epsilon_{ij}\epsilon_{kl}
\langle v^{0}|(\partial_k\partial_i\widehat H)|
\partial_j\partial_l v^{0}\rangle
\nonumber\\
&-\frac{B^{2}}{4}\,\epsilon_{ij}\epsilon_{kl}
\langle\partial_i v^{0}|(\partial_j\partial_k\widehat H)|
\partial_l v^{0}\rangle
\nonumber\\
&-\frac{B^{2}}{4}\,\epsilon_{ij}\epsilon_{kl}
\langle\partial_i v^{0}|(\partial_k\widehat H)|
\partial_j\partial_l v^{0}\rangle
\nonumber\\
&-B\,\epsilon_{ij}\,
\mathrm{Im}\langle v^{1}|(\partial_i\widehat H)|\partial_j v^{0}\rangle
+\langle v^{1}|(\widehat H-\varepsilon_0)|v^{1}\rangle .
\label{eq:secondraw}
\end{align}

The last two terms need the first-order state, which is determined
in two independent pieces. The normalization
$\langle v|\!\star\!|v\rangle=1$ at $O(B)$ fixes the component of
$|v^{1}\rangle$ along $|v^{0}\rangle$,
\begin{equation}
2\,\mathrm{Re}\langle v^{0}|v^{1}\rangle
+\frac{iB}{2}\,\epsilon_{ij}
\langle\partial_i v^{0}|\partial_j v^{0}\rangle = 0
\;\Rightarrow\;
\mathrm{Re}\langle v^{0}|v^{1}\rangle=\frac{B\Omega}{4},
\label{eq:norm1}
\end{equation}
where $\epsilon_{ij}\langle\partial_i v^{0}|\partial_j
v^{0}\rangle=i\Omega$ is the Berry curvature in our conventions.
The imaginary part is left free. The interband part comes from the
eigenvalue equation \eqref{eq:star} at $O(B)$,
\begin{equation}
(\widehat H-\varepsilon_0)|v^{1}\rangle
= B\,\morb\,|v^{0}\rangle
-\frac{iB}{2}\,\epsilon_{ij}\,(\partial_i\widehat H)|\partial_j v^{0}\rangle,
\label{eq:v1eq}
\end{equation}
which, projected onto a remote band $\langle b^{0}|$, yields the
coefficient of $|b^{0}\rangle$. In the band basis
$\{|b^{0}\rangle\}$,
\begin{equation}
|v^{1}\rangle = B\Big(\frac{\Omega}{4}+i\alpha\Big)|v^{0}\rangle
-\frac{iB}{2}\,\epsilon_{ij}\sum_{b\neq v}
\frac{\langle b^{0}|\partial_i\widehat H|\partial_j v^{0}\rangle}
{\varepsilon_b-\varepsilon_0}\,|b^{0}\rangle,
\label{eq:v1sol}
\end{equation}
with $\alpha$ the undetermined real phase left by the
normalization.

Projecting Eq.~\eqref{eq:v1eq} onto $\langle v^{1}|$ and taking the real part, with Eq.~\eqref{eq:norm1} supplying
$\mathrm{Re}\langle v^{1}|v^{0}\rangle$, converts the $\mathrm{Im}$ term of Eq.~\eqref{eq:secondraw} into
\begin{equation}
-B\,\epsilon_{ij}\,
\mathrm{Im}\langle v^{1}|(\partial_i\widehat H)|\partial_j v^{0}\rangle
= \frac{B^{2}\morb\,\Omega}{2}
- 2\,\langle v^{1}|(\widehat H-\varepsilon_0)|v^{1}\rangle ,
\label{eq:imred}
\end{equation}
while the $\langle v^{1}|(\widehat H-\varepsilon_0)|v^{1}\rangle$
term follows from Eq.~\eqref{eq:v1sol} directly. Only the interband part survives the factor
$\widehat H-\varepsilon_0$, and $\alpha$ drops out,
\begin{align}
&\langle v^{1}|(\widehat H-\varepsilon_0)|v^{1}\rangle
\nonumber\\
&\quad= \frac{B^{2}}{4}\,\epsilon_{ij}\epsilon_{kl}\sum_{b\neq v}
\frac{\langle\partial_l v^{0}|(\partial_k\widehat H)|b^{0}\rangle
\langle b^{0}|(\partial_i\widehat H)|\partial_j v^{0}\rangle}
{\varepsilon_b-\varepsilon_0} .
\label{eq:quad}
\end{align}
Both terms are now free of $\alpha$. This phase, which is the
imaginary part of $\langle v^{0}|v^{1}\rangle$ up to the factor
$B$, never enters the flat-band dispersion.

One can further simplify the first term of Eq.~\eqref{eq:secondraw}. Differentiating Eq.~\eqref{eq:id1} once more gives
\begin{align}
(\partial_k\partial_i\widehat H)|v^{0}\rangle
={}& -(\partial_k\widehat H)|\partial_i v^{0}\rangle
-(\partial_i\widehat H)|\partial_k v^{0}\rangle
\nonumber\\
&-(\widehat H-\varepsilon_0)|\partial_k\partial_i v^{0}\rangle ,
\label{eq:id2}
\end{align}
which converts it into matrix elements with single derivatives. Under contraction with the Levi-Civita symbols
$\epsilon_{ij}\epsilon_{kl}$, one of the resulting pieces cancels the
third term of Eq.~\eqref{eq:secondraw} exactly. Collecting
Eqs.~\eqref{eq:secondraw}--\eqref{eq:id2},
$\varepsilon^{(2)}=\tfrac12 B^{2}\cchi_{\rm raw}$ with the
closed form
\begin{align}
\cchi_{\rm raw} ={}& \morb\,\Omega
-\frac{1}{2}\,\epsilon_{ij}\epsilon_{kl}
\langle\partial_i v^{0}|(\partial_j\partial_k\widehat H)|
\partial_l v^{0}\rangle
\nonumber\\
&+\frac{1}{4}\,\epsilon_{ij}\epsilon_{kl}
\langle\partial_k\partial_i v^{0}|(\widehat H-\varepsilon_0)|
\partial_j\partial_l v^{0}\rangle
\nonumber\\
&-\frac{1}{2}\,\epsilon_{ij}\epsilon_{kl}\sum_{b\neq v}
\frac{\langle\partial_l v^{0}|(\partial_k\widehat H)|b^{0}\rangle
\langle b^{0}|(\partial_i\widehat H)|\partial_j v^{0}\rangle}
{\varepsilon_b-\varepsilon_0}.
\label{eq:chidef}
\end{align}
Every object on the right-hand side is built from the zero-field
wavefunctions. The first term couples the moment to the Berry
curvature. The middle terms carry second derivatives of the
Hamiltonian and of the band projector. The last is a cross-gap sum
over the remote bands.

Unlike the moment, the second-order coefficient is not gauge
invariant by itself. This is the order at which the Berry
connection $\mathcal A_i = i\langle v^{0}|\partial_i v^{0}\rangle$
first enters. The dependence follows from a covariance of the
star-eigenvalue problem. A change of phase convention of the zero-field state,
$|v^{0}\rangle \to e^{i\phi(\bm\pi)}|v^{0}\rangle$, is realized on
the full solution of Eq.~\eqref{eq:star} as a star-multiplication
from the right, $|v\rangle\to|v\rangle\star e_\star^{i\phi}$, where
$e_\star^{i\phi}=1+i\phi+\tfrac{1}{2}(i\phi)\star(i\phi)+\cdots$ is
the star-exponential, with inverse $e_\star^{-i\phi}$. The transformed
state solves Eq.~\eqref{eq:star} with the conjugated eigenvalue
$\varepsilon \to e_\star^{-i\phi}\star\varepsilon\star e_\star^{i\phi}$
\cite{YeWang2025},
which to the retained order is a translation of the symbol,
\begin{equation}
\varepsilon(\bm\pi) \;\to\;
\varepsilon\big(\pi_i - B\,\epsilon_{ij}\,\partial_j\phi\big)
+ O(B^{3}) ,
\label{eq:cov}
\end{equation}
the correction being $O(B^{3})$ rather than $O(B^{2})$ because
$\varepsilon_0$ is constant. Applied to
$\varepsilon=\varepsilon_0+B\,\morb(\bm\pi)
+\tfrac12 B^{2}\cchi_{\rm raw}(\bm\pi)$, the translation acts only
through the first-order term,
\begin{align}
\varepsilon(\bm\pi) \;\to\; \varepsilon_0 + B\,\morb(\bm\pi)
- &B^{2}\epsilon_{ij}\big(\partial_i\morb\big)\partial_j\phi
\nonumber\\&+ \tfrac12 B^{2}\cchi_{\rm raw}(\bm\pi) + O(B^{3}) .
\label{eq:covexpand}
\end{align}
The moment is gauge invariant and the connection transforms as
$\mathcal A_j\to\mathcal A_j-\partial_j\phi$, so the single
combination $B^{2}\epsilon_{ij}(\partial_i\morb)\mathcal A_j$
reproduces the entire variation in Eq.~\eqref{eq:covexpand}.
Splitting it off from the second-order term,
\begin{equation}
\cchi(\bm\pi) \;\equiv\; \cchi_{\rm raw}(\bm\pi)
\;-\; 2\,\epsilon_{ij}\,\big(\partial_i\morb\big)\,\mathcal A_j ,
\label{eq:chisplit}
\end{equation}
leaves a remainder that must be unchanged by the gauge
transformation: $\cchi$ is gauge invariant.%
\footnote{The gauge invariance uses $\partial\varepsilon_0=0$. For a
dispersive band the same substitution removes the connection from the
first-order term,
$\varepsilon_0(\bm\pi)+B\epsilon_{ij}(\partial_i\varepsilon_0)\mathcal A_j
=\varepsilon_0(\mathbf k)+O(B^{2})$ \cite{YeWang2025}, but the
second-order term then contains
$-\epsilon_{ij}(\partial_i\varepsilon_0)\,\partial_j\alpha$, with
$\alpha$ the phase of Eq.~\eqref{eq:v1sol}, which no function of the
zero-field connection can absorb. Here $\alpha$ drops out,
Eq.~\eqref{eq:quad}.}

The split has a compact form, since the connection term is the
first-order Taylor term of $B\,\morb$ at a shifted argument. Defining
the shifted momentum of Ref.~\cite{YeWang2025},
\begin{equation}
k_i \;=\; \pi_i + B\,\epsilon_{ij}\,\mathcal A_j ,
\label{eq:kmod}
\end{equation}
absorbs it, and with $\cchi(\mathbf k)=\cchi(\bm\pi)+O(B)$ the
dispersion reads
\begin{equation}
\morb(\bm\pi)B + \frac{1}{2}\cchi_{\rm raw}(\bm\pi)B^{2}
= \morb(\mathbf k)B + \frac{1}{2}\cchi(\mathbf k)B^{2}
+ O(B^{3}) .
\label{eq:absorb}
\end{equation}
This is also what makes the invariance of observables manifest.
Every physical quantity is a phase-space integral, and
Eq.~\eqref{eq:kmod} leaves behind only its Jacobian,
$d^{2}\pi=(1+B\,\Omega)\,d^{2}k$ [Eq.~\eqref{eq:Xidef}]. Integrand
and measure are then separately gauge invariant, and the gauge
dependence of the integration variable, an area-preserving shift with
$\epsilon_{ij}\partial_i\partial_j\phi=0$, drops out.

The closed form of $\cchi$ can be shown to be Eq.~\eqref{eq:chidef}
with every bare derivative of the band state replaced by its
covariant counterpart $D_i = \partial_i + i\mathcal A_i$:
\begin{align}
\cchi ={}& \morb\,\Omega
-\frac{1}{2}\,\epsilon_{ij}\epsilon_{kl}
\langle D_i v^{0}|(\partial_j\partial_k\widehat H)|
D_l v^{0}\rangle
\nonumber\\
&+\frac{1}{4}\,\epsilon_{ij}\epsilon_{kl}
\langle D_k D_i v^{0}|(\widehat H-\varepsilon_0)|
D_j D_l v^{0}\rangle
\nonumber\\
&-\frac{1}{2}\,\epsilon_{ij}\epsilon_{kl}\sum_{b\neq v}
\frac{\langle D_l v^{0}|(\partial_k\widehat H)|b^{0}\rangle
\langle b^{0}|(\partial_i\widehat H)|D_j v^{0}\rangle}
{\varepsilon_b-\varepsilon_0} ,
\label{eq:chifull}
\end{align}
which is gauge invariant term by term. In the numerical checks of
Sec.~\ref{sec:numerics}, $\cchi$ is evaluated pointwise in the parallel-transport gauge, $\mathcal A=0$ at the
evaluation point, in which Eq.~\eqref{eq:chifull} reduces to
Eq.~\eqref{eq:chidef}, with the derivatives of $|v^{0}\rangle$ and
$\widehat H$ taken by finite differences.

The Jacobian of the substitution \eqref{eq:kmod}, $1+B\,\Omega$, is a
direct consequence of gauge invariance under the Moyal product. The
same phase-space volume factor is known from the semiclassical
approximation \cite{SundaramNiu1999,XiaoShiNiu2005,Xiao2010}, and
here it applies at $O(B^{2})$ and for an exactly flat band. The
assembled dispersion is then gauge invariant term by term:
\begin{equation}
\varepsilon(\mathbf k;B)=\varepsilon_0+\morb(\mathbf k)\,B
+\frac{1}{2}\,\cchi(\mathbf k)\,B^{2}+O(B^{3}).
\label{eq:compact}
\end{equation}

The second-order dispersion of Ref.~\cite{PHSB2026}, obtained from
the same Moyal expansion, reduces on an exactly flat band and in the
present notation to
\begin{equation}
\cchi_{\rm PHSB} \;=\; \cchi \;-\; \morb\,\Omega
\;+\; \frac{1}{2}\,\varepsilon_0\,\Omega^{2} .
\label{eq:phsb}
\end{equation}
The last term violates gauge invariance under time-dependent gauge
transformations. A gauge function linear in time leaves the vector
potential unchanged and shifts the scalar potential, and with it
$\widehat H$ and every band energy, by a constant $c$. Under this shift
of the energy zero, $\widehat H\to\widehat H+c$, the last term changes by $\tfrac12 c\,\Omega^{2}$,
whose zone integral is nonzero for any band with nonvanishing Berry
curvature, although no energy difference in the problem is altered.
Every term of Eq.~\eqref{eq:chifull} is unchanged, since the
Hamiltonian enters only through $\widehat H-\varepsilon_0$ and its
derivatives.

\subsection{The expansion parameter}
\label{sec:control}

For a dispersive band the ratio $\hbar\omega_c/E_{\rm F}$ of the
cyclotron energy to the Fermi energy is an obvious small parameter in the semiclassical expansion. A flat band has neither scale; the effective mass diverges, so the cyclotron energy
$\hbar\omega_c\propto B/m^{*}$ vanishes at leading order, and the
Fermi energy measured from the band is zero. The expansion parameter is instead set by the Moyal product.

Every order of the star product \eqref{eq:star} inserts one factor
of $B$ and one derivative pair $\epsilon_{ij}\partial_i\partial_j$,
so the expansion runs formally in $B(\partial_k)^{2}$, the field times the square of a length, whose value is set by the momentum dependence on which the derivatives act. On a flat band
$\varepsilon_0$ has none, so every derivative lands on the
wavefunctions, and their momentum dependence is measured by the
quantum metric
$g_{ij}=\mathrm{Re}\,\langle\partial_i v^{0}|Q|\partial_j
v^{0}\rangle$, with $Q=1-|v^{0}\rangle\langle v^{0}|$. We define the
geometric length $\xi$ of the band by
$\xi^{2}=4\max_{\mathbf k}\lambda_{\max}(g)$, where
$\lambda_{\max}(g)$ is the larger eigenvalue of $g$. The factor of
four calibrates $\xi$ to the length $v/|m|$ of a massive Dirac cone
with velocity $v$ and half-gap $|m|$, evaluated at the end of this
subsection. Every projected derivative obeys
$\|Q|\partial_i v^{0}\rangle\|\le\xi$. Each
wavefunction derivative in the expansion therefore contributes a
factor of at most $\xi$, and the dimensionless expansion parameter
is
\begin{equation}
x \;=\; B\,\xi^{2} \;=\; \Big(\frac{\xi}{\ell_B}\Big)^{2}\;\ll\;1 ,
\label{eq:control}
\end{equation}
the flux through the geometric area $\xi^{2}$, equivalently the
squared ratio of $\xi$ to the magnetic length $\ell_B=B^{-1/2}$.

The length $\xi$ has a direct physical reading. For a band that
admits exponentially localized Wannier functions, the zone average
of $\mathrm{tr}\,g$ is the gauge-invariant part of the quadratic
spread of the maximally localized Wannier function
\cite{MarzariVanderbilt1997}. Thus $\xi$ measures the linear size
of the most localized orbital the band supports, and $x$ is the
flux through that orbital.

The closed forms verify this counting. Equation~\eqref{eq:morb}
carries two projected derivatives (the $(\widehat H-\varepsilon_0)$ insertion retains only
$Q|\partial_i v^{0}\rangle$, each bounded by $\xi$) and one gap-scale insertion, and every term of Eq.~\eqref{eq:chidef} carries
two more wavefunction derivatives, each contributing a further
factor of $\xi$. Hence
\begin{equation}
|\morb| \;\sim\; \Delta\,\xi^{2},
\quad
|\cchi| \;\sim\; \Delta\,\xi^{4}
\quad\Longrightarrow\quad
\frac{\frac{1}{2}\,\cchi\,B^{2}}{\morb\,B}
\;\sim\; B\,\xi^{2},
\label{eq:mscale}
\end{equation}
and the $n$-th Moyal order is of size $\Delta(B\xi^{2})^{n}$.%
\footnote{This estimate uses a single energy scale $\Delta$. The gap
appears directly only in $\morb$ and in the energy denominator of the
cross-gap term of Eq.~\eqref{eq:chidef}, dominated by the nearest
remote band. That term and the band-curvature term
$\partial_j\partial_k\widehat H$ also involve the velocities of the
remote sector, $\langle b^{0}|\partial_i\widehat H|c^{0}\rangle$,
which the flat-band metric does not constrain, and we take them to be
of the same scale, $\sim\Delta\xi$. If the nearest remote band is
anomalously dispersive, its velocity scale replaces $\Delta\xi$ there
and enlarges the effective expansion parameter accordingly.}
Near the gap minimum of a massive Dirac cone
with velocity $v$ and half-gap $|m|$, the length defined above is
$\xi=v/|m|$, and the scaling \eqref{eq:mscale} is realized exactly,
\begin{equation}
\morb \;=\; -\frac{1}{2}\,|m|\,\xi^{2} ,
\label{eq:new-diraccal}
\end{equation}
with the half-gap in place of $\Delta$.

\section{The orbital magnetic susceptibility of a flat band}
\label{sec:chi}

At the noninteracting level, the susceptibility of a flat band is
contained in the gauge-invariant multiband formulas
\cite{Raoux2015,GaoNiu2017} as the limit of vanishing dispersion. This section derives it along a different
route, from the field-generated dispersion of
Sec.~\ref{sec:dispersion}, and proves that the two routes agree.
Interacting extensions exist.
Ref.~\cite{Ye2026} formulates the interacting problem in the
noncommutative kinetic momentum, the algebra of
Sec.~\ref{sec:dispersion}, and obtains the orbital magnetization at
first order in $B$. Ref.~\cite{KWV2025}, working from the grand
potential at finite field, carries the interacting susceptibility
to second order at self-consistent mean-field level, as a band
functional supplemented by the feedback of the field-dependent mean
field. Ref.~\cite{MayrhoferChubukov2023} shows that interactions
render the Landau diamagnetic susceptibility of a two-dimensional
Fermi liquid nonanalytic in temperature and in-plane field.

The grand potential per unit area is the phase-space trace of
$\FT(\widehat H)$, the trace per unit area being
$\int\!d^{2}\pi/(2\pi)^{2}$ of the symbol \cite{Roth1962}, with
$\FT(\varepsilon)=-T\ln\!\big(1+e^{-(\varepsilon-\mu)/T}\big)$.
Two steps convert the trace to an integral. First, the trace of a star product of two symbols is the plain
integral of their ordinary product,
\begin{equation}
\int\!\frac{d^{2}\pi}{(2\pi)^{2}}\; a\star b
=\int\!\frac{d^{2}\pi}{(2\pi)^{2}}\; a\,b ,
\label{eq:new-startrace}
\end{equation}
because every insertion of the star exponential
\eqref{eq:star} beyond the zeroth is a total derivative under the
integral. Second, the symbol of $\FT(\widehat H)$ is
$\FT(\varepsilon)$ up to corrections from the star products among
the factors of $\varepsilon$ in the power series of $\FT$. In the star product \eqref{eq:star} every power of $B$ comes with one
derivative acting on each of the two factors, so each correction
differentiates at least two factors of $\varepsilon$. A flat band has no zero-field
dispersion, so a differentiated factor of $\varepsilon$ is itself
$O(B)$. The lowest correction, one explicit power of $B$ times two such
factors, is therefore $O(B^{3})$, beyond the order retained. With the
dispersion of Sec.~\ref{sec:second},
\begin{equation}
\Xi(B) = \int\!\frac{d^{2}\pi}{(2\pi)^{2}}
\FT\big(\varepsilon_0 + \morb(\bm\pi)B
+ \frac{1}{2}\cchi_{\rm raw}(\bm\pi)B^{2}\big) \!+ O(B^{3}),
\label{eq:Xipi}
\end{equation}
with flat measure but the gauge-dependent coefficient. The identity \eqref{eq:absorb} replaces
$\cchi_{\rm raw}(\bm\pi)$ by $\cchi(\mathbf k)$ in the argument,
and its Jacobian, the Berry-measure factor $1+B\,\Omega$, affects
the measure,
\begin{equation}
\Xi(B) = \int\!\frac{d^{2}k}{(2\pi)^{2}}\,
\big(1 + B\,\Omega\big)\,\FT\big(\varepsilon(\mathbf k;B)\big) .
\label{eq:Xidef}
\end{equation}
Every object in Eq.~\eqref{eq:Xidef} is gauge invariant.
Differentiating at $B=0$ gives the magnetization and susceptibility,
\begin{align}
M &= -\partial_B \Xi\big|_{0}
= -\!\int\!\frac{d^{2}k}{(2\pi)^{2}}
\big[\morb\, f + \Omega\,\FT\big]_{\varepsilon_0},
\label{eq:Mthermo}\\
\chi_{\rm thermo} &= -\partial_B^{2}\,\Xi\big|_{0}
\nonumber\\
&= -\!\int\!\frac{d^{2}k}{(2\pi)^{2}}
\big[(\cchi + 2\morb\Omega)\,f
+ \morb^{2}\,f'\big]_{\varepsilon_0},
\label{eq:chithermo}
\end{align}
with $f=\FT'$ the Fermi function and all integrands evaluated on
the flat band. The magnetization is
the thermally weighted moment plus the Berry-measure correction, and
the susceptibility collects $\cchi$, the moment--curvature cross
term, and the Curie-like $\morb^{2}f'$ term of the field-generated
moments.

Equation~\eqref{eq:chithermo} agrees with the gauge-invariant
susceptibility of Ref.~\cite{Raoux2015} applied to a flat band: the
exact trace formula
\begin{align}
\chi_{\rm thermo} ={}& -\frac{1}{12}\,\frac{\mathrm{Im}}{\pi}
\int\! dE\; f(E) \int\!\frac{d^{2}k}{(2\pi)^{2}}\,
\operatorname{Tr}\big\{ G h^{xx} G h^{yy}
\nonumber\\
& - G h^{xy} G h^{xy}
- 4\big(G h^x G h^x G h^y G h^y
\nonumber\\
&\hspace{5.2em} - G h^x G h^y G h^x G h^y\big)\big\},
\label{eq:master}
\end{align}
with $G=(E-h+i0)^{-1}$,
$h^{a}\equiv\partial h/\partial k_{a}$, and
$h^{ab}\equiv\partial^{2}h/\partial k_{a}\partial k_{b}$, where
$h(\mathbf k)$ is the zero-field Bloch Hamiltonian of
Sec.~\ref{sec:moyal}, here a function of the crystal momentum, and
the indices $a,b$ run over $x,y$.
In Appendix~\ref{app:consistency} we reduce
Eq.~\eqref{eq:master} at an arbitrary isolated, nondegenerate flat
band and recover the structure and the coefficients of
Eq.~\eqref{eq:chithermo}: the same
two thermal weights $f(\varepsilon_0)$ and $f'(\varepsilon_0)$ and
no others, a Curie coefficient that matches $-\morb^{2}$ pointwise
for any number of bands, and an $f$-coefficient that matches
$-[\cchi+2\morb\Omega]$ after zone integration. The susceptibility
of an isolated flat band is thus fixed by three functions of the zero-field wavefunctions ($\morb$,
$\cchi$, $\Omega$), in agreement with the established
gauge-invariant response theory \cite{Raoux2015,GaoNiu2017,KWV2025}.

The procedure of this section is not restricted to the flat band.
The same two steps, expand the dispersion in the kinetic momentum
and differentiate the grand potential, apply to a dispersive band,
where $\varepsilon_0(\bm{\pi})$ is momentum dependent in the argument of $\FT$ \cite{HueckerYeWang_prep}. They apply also to
interacting systems treated at self-consistent mean-field level,
where the expansion runs over the Hartree--Fock bands and the field
dependence of the mean field enters as a feedback term
\cite{KWV2025,Ye2026}.

For a two-band model the sums over remote bands can be carried out
in closed form, and we record the result here for later use in
Sec.~\ref{sec:numerics}. Evaluating the cross-gap sum of Eq.~\eqref{eq:chidef}, which for
the two-band problem runs over a single remote band, gives
\begin{equation}
\cchi(\mathbf k) = \frac{1}{4}\,U_2(\mathbf k) - \Omega^{2}(\mathbf k),
\qquad
U_2 \equiv \hat{\boldsymbol d}_{xx}\!\cdot\hat{\boldsymbol d}_{yy}
- \hat{\boldsymbol d}_{xy}\!\cdot\hat{\boldsymbol d}_{xy},
\label{eq:chi2band}
\end{equation}
where $\hat{\boldsymbol d}$ is the unit Bloch vector of the model,
$\widehat H=\hat{\boldsymbol d}\cdot\bm\sigma$, and subscripts denote
momentum derivatives. $U_2$ is a total divergence, so its zone
integral vanishes. The full susceptibility then follows from the trace-formula
reduction of Appendix~\ref{app:consistency}. The remote-band
(frozen) contribution vanishes, because the resolvent has no bands
beyond the pair whose pole terms are kept, and assembling the two
bands ($\varepsilon\to-\varepsilon$ sends $\Omega\to-\Omega$ and
$\cchi\to-\cchi$ with $\morb$ unchanged) gives
\begin{align}
\chi_{\rm thermo}(\mu,T) &= -\,I\,
\big[f(-1)-f(+1)+f'(-1)+f'(+1)\big],
\nonumber\\
I &\equiv \int\!\frac{d^{2}k}{(2\pi)^{2}}\,\Omega^{2},
\label{eq:chiclosedform}
\end{align}
a single wavefunction integral carrying the full
$(\mu,T)$-dependence in the bracket of Fermi factors.

\section{The de Haas--van Alphen effect of a flat band}
\label{sec:LK}

The response of Sec.~\ref{sec:chi} is smooth in $B$ and therefore
contains no de Haas--van Alphen oscillations. The reason lies in a single step. Converting the trace of $\FT(\widehat H)$ into the
kinetic-momentum integral \eqref{eq:Xipi} replaced a sum over the
discrete spectrum of the star-eigenvalue problem by the integral of
a smooth symbol, order by order in $B$. The discreteness of the
spectrum survives in that conversion only through corrections
nonanalytic in $B$, which vanish at every order of the expansion. One route
to the oscillations is therefore to redo the conversion carefully,
keeping the level structure of the star spectrum and restoring the
nonanalytic terms by Poisson summation
\cite{LK1956,Shoenberg1984,Roth1966,Fuchs2018}. There is an alternative.
At fixed chemical potential the oscillations are carried entirely by
the low-energy excitations near $\mu$, and for a Fermi system those
excitations admit an exact bosonized description. We take the second
route.

For the flat band the low-energy structure is generated by the field
itself. Pin $\mu=\varepsilon_0$ and let $B>0$. States with
$\morb<0$ are filled, states with $\morb>0$ are empty, and the
Fermi surface is the zero contour $\mathcal C_0$ of
Eq.~\eqref{eq:zerocontour}. We assume $\morb$ changes sign and
$\mathcal C_0$ exists. Ref.~\cite{HwangRhimYang2021} showed that
the Landau levels of an isolated flat band, with spacing
$\sim B^{2}$, are those of the effective dispersion
$B\,\morb(\mathbf k)$, the first-order dispersion of
Sec.~\ref{sec:first}. One thus expects quantum oscillations.

It is not guaranteed that the semiclassical rules of a dispersive
band apply to an emergent Fermi surface. In a metal the
quantization phase is the Maslov one-half plus the Berry phase of a
zero-field orbit. Here the orbit itself exists only at finite field,
its dispersion carries the second-order coefficient $\cchi$ alongside
the moment, and how the phase is related to the Berry curvature and
to these other terms of the field-generated dispersion is not
settled by analogy. The derivation below decides it without
semiclassical input. No quantization rule is assumed, and the phase
comes out of the topology of the bosonized theory. It holds for an
isolated, nondegenerate flat band whose moment $\morb(\mathbf k)$
is smooth with a single closed $\morb=0$ contour.

\subsection{Bosonized action and projection to the flat band}
\label{sec:boson}

Our derivation uses coadjoint-orbit bosonization, proposed in
Ref.~\cite{DelacretazDuMehtaSon2022} as an exact bosonic
description of a Fermi surface in terms of its shape fluctuations.
It was applied to the de Haas--van Alphen effect of a Fermi gas
and a Fermi liquid in Ref.~\cite{Wang2025} and extended to
multiband systems in Ref.~\cite{YeWang2025}. The multiband version
is the method we apply to the flat band. For a single Fermi contour
the fluctuations organize into one chiral boson $\phi(\theta,t)$,
with $\theta$ a coordinate along the contour, defined in
Sec.~\ref{sec:contour}. Because the boson is compact, its
configuration space splits into topological sectors labelled by an
integer winding number, which physically represents the number of particles added to the contour.

The action of the multiband construction, obtained in
Ref.~\cite{YeWang2025} from the fermion path integral over
particle--hole coherent states, is
\begin{equation}
S=\int dt\,\frac{d\mathbf x\,d\mathbf p}{(2\pi)^{2}}\,
\mathrm{Tr}\Big[\widehat f_0\star\widehat U^{-1}\star i\partial_t\widehat U
-\widehat f\star\widehat H\Big],
\label{eq:boson}
\end{equation}
where every hatted object is a matrix in the band (orbital) space
and $\mathrm{Tr}$ traces over that space. The reference
distribution $\widehat f_0$ is the Wigner transform of the
ground-state one-particle density matrix
$\rho_0(\mathbf x,\mathbf x')$,
\begin{equation}
\widehat f_0(\mathbf x,\mathbf p)
=\int d^{2}s\;e^{-i\mathbf p\cdot\mathbf s}\,
\rho_0\Big(\mathbf x+\frac{\mathbf s}{2},
\mathbf x-\frac{\mathbf s}{2}\Big),
\label{eq:new-wigner}
\end{equation}
a matrix in the orbital indices. The unitary
$\widehat U=e^{i\widehat\phi}$, with
$\widehat\phi(\mathbf x,\mathbf p)$ a Hermitian matrix symbol,
generates the particle--hole rotations of the ground state, and
$\widehat f=\widehat U\star\widehat f_0\star\widehat U^{-1}$ is the
rotated distribution. The $\star$ in Eq.~\eqref{eq:boson} is
the phase-space Moyal product
\begin{equation}
\star=\exp\Big[\frac{i}{2}\big(
\overleftarrow{\partial}_{\mathbf x}\!\cdot\overrightarrow{\partial}_{\mathbf p}
-\overleftarrow{\partial}_{\mathbf p}\!\cdot\overrightarrow{\partial}_{\mathbf x}
\big)\Big],
\label{eq:starxp}
\end{equation}
defined by gradients in $\mathbf x$ and $\mathbf p$. It is not the
momentum-space product of Eq.~\eqref{eq:star}, which acts on
functions of $\bm\pi$ alone. No expansion has been made. The action
\eqref{eq:boson} reproduces the full fermion partition function
restricted to the particle--hole coherent states
$\widehat U|G\rangle$, with $|G\rangle$ the ground state
\cite{YeWang2025}.

Two changes of variables bring the action to the flat band
\cite{YeWangPRB2025,YeWang2025}. The first is
$(\mathbf x,\mathbf p)\to(\mathbf R,\bm\pi)$, with
$\bm\pi=\mathbf p-\mathbf A(\mathbf x)$ the kinetic momentum
($\mathbf A$ is the electromagnetic vector potential) and
$R_i=x_i+\epsilon_{ij}\pi_j/B$ the guiding center. The two sets
$\star$-commute, and the product \eqref{eq:starxp} factorizes into
a $\bm\pi$-sector product, which is Eq.~\eqref{eq:star}, and an
$\mathbf R$-sector product of the same form with $B\to-1/B$, where for
$[f,g]_\star\equiv f\star g-g\star f$,
\begin{equation}
[\pi_i,\pi_j]_\star=iB\,\epsilon_{ij},\quad
[R_i,R_j]_\star=-\frac{i}{B}\,\epsilon_{ij},\quad
[R_i,\pi_j]_\star=0
\label{eq:RPialg}
\end{equation}
\cite{YeWangPRB2025}. We project onto the flat band with the same
$\star$-eigenstates that built Eq.~\eqref{eq:compact}. Every matrix
symbol is expanded in the star-eigenbasis of the bands, and the
flat-band diagonal element is kept,
\begin{equation}
f_0(\mathbf k)=\langle v|\!\star\!\widehat f_0\!\star\!|v\rangle,
\qquad
\phi(\mathbf R,\mathbf k)=\langle v|\!\star\!\widehat\phi\!\star\!|v\rangle .
\label{eq:new-project}
\end{equation}
The off-band elements of $\widehat\phi$ create particle--hole pairs
across the gap. Their contribution to the action carries the energy
denominator $\Delta$, while every retained scale is a power of the
field, so they enter beyond the retained order and the flat-band
scalars \eqref{eq:new-project} are the fluid variables. The second
change is $(\mathbf R,\bm\pi)\to(\mathbf R,\mathbf k)$, by the
shifted momentum of Eq.~\eqref{eq:kmod} \cite{YeWang2025}.
In the conventions of Sec.~\ref{sec:second} the connection obeys
$\epsilon_{ij}\partial_i\mathcal A_j=-\Omega$, so the Jacobian of
Eq.~\eqref{eq:kmod} is
$\det(\partial k/\partial\pi)=1+B\,\epsilon_{ij}\partial_i\mathcal
A_j=1-B\Omega$, and the measure transforms as
$d^{2}\pi=(1+B\Omega)\,d^{2}k$ to the retained order, the measure
of Eq.~\eqref{eq:Xidef}. The momentum-sector bracket carries the
inverse factor, using $[\pi_i,f]_\star=iB\,\epsilon_{il}\partial_l f$,
\begin{align}
[k_i,k_j]_\star
&=[\pi_i,\pi_j]_\star
+B\,\epsilon_{jm}[\pi_i,\mathcal A_m]_\star
+B\,\epsilon_{il}[\mathcal A_l,\pi_j]_\star
\nonumber\\
&=iB\,\epsilon_{ij}\big(1+B\,\epsilon_{lm}\partial_l\mathcal A_m\big)
=\frac{iB\,\epsilon_{ij}}{1+B\Omega}
\label{eq:new-kbracket}
\end{align}
to the retained order. The result is a single-band fluid
\cite{YeWang2025}:
\begin{equation}
S=\int_{t,\mathbf R,\mathbf k}\!\!(1+B\Omega)
\Big[f_0(\mathbf k)\,U^{-1}\star i\partial_t U
- f\star\varepsilon(\mathbf k;B)\Big],
\label{eq:fluid}
\end{equation}
where $f_0=\Theta(\mu-\varepsilon)$, with $\Theta$ the Heaviside
step function, is the zero-temperature value of the projected
occupation \eqref{eq:new-project}, and $\star$ is the product in
the $(\mathbf R,\mathbf k)$ variables. One ingredient extends
Ref.~\cite{YeWang2025}. The dispersion entering
Eq.~\eqref{eq:fluid} is the full second-order symbol of
Eq.~\eqref{eq:compact}. The analysis of Ref.~\cite{YeWang2025} stops
at first order. For the flat band the second-order term survives in
the phase.

\subsection{Chiral boson on the emergent contour}
\label{sec:contour}

The reduction to the contour proceeds in two steps. Magnetic
translation symmetry reduces the two-dimensional boson to
$N_\Phi$ copies, and the expansion around the ground state
localizes each copy on the emergent Fermi contour.

By magnetic translation symmetry it suffices to keep configurations
invariant under magnetic translations \cite{YeWang2025}, the
symmetry-breaking configurations contributing only Gaussian factors
smooth in $B$, and the invariant ones carry no dependence on the
guiding center. The
guiding-center plane is not a passive label. By the algebra
\eqref{eq:RPialg} it is a noncommutative plane carrying
$N_\Phi=BL^{2}/2\pi$ states, with $L^{2}$ the system area, so the
$\mathbf R$-integral in
Eq.~\eqref{eq:fluid} is a trace over those states, and an
$\mathbf R$-independent configuration contributes one identical
copy per state. Thus $\phi(\mathbf R,\mathbf k)$ reduces to $N_\Phi$
copies $\phi_i(\mathbf k)$, one per magnetic state.
Recall that $\phi$ enters the action through $U=e^{i\phi}$, so an
expansion in $\phi_i$ is an expansion of Eq.~\eqref{eq:fluid}
around the ground state. To quadratic order the two factors of
Eq.~\eqref{eq:fluid} expand as
\begin{align}
U^{-1}\star i\partial_t U
&= -\dot\phi_i+\frac{i}{2}\,[\phi_i,\dot\phi_i]_\star+O(\phi_i^{3}),
\nonumber\\
f &= f_0+i\,[\phi_i,f_0]_\star
-\frac{1}{2}\,\big[\phi_i,[\phi_i,f_0]_\star\big]_\star+O(\phi_i^{3}),
\label{eq:new-quadratic}
\end{align}
where, by the bracket \eqref{eq:new-kbracket}, to the retained
order
$[a,b]_\star=\tfrac{iB}{1+B\Omega}\,\epsilon_{lm}\,
\partial_l a\,\partial_m b$, with $l,m$ running over $k_x,k_y$.
For nonwinding configurations the terms linear in $\phi_i$
integrate to zero. The ground state extremizes the action. The
winding sector is kept below. Two structures organize the quadratic
terms. First, every single commutator with $f_0$ is supported on
the emergent contour, where $f_0$ jumps:
\begin{equation}
[\phi_i,f_0]_\star
=-\frac{iB}{1+B\Omega}\,
\delta\big(\varepsilon(\mathbf k;B)-\mu\big)\,
\epsilon_{lm}\,\partial_l\phi_i\,\partial_m\varepsilon ,
\label{eq:new-localize}
\end{equation}
and $\epsilon_{lm}\,\partial_l\phi_i\,\partial_m\varepsilon$ is
$|\partial_{\mathbf k}\varepsilon|/|\partial_\theta\mathbf k|$
times the derivative of $\phi_i$ along the contour, where $\theta$
parametrizes the contour counterclockwise, $\mathbf k(\theta)$.
Terms carrying this support involve only the contour modes
$\phi_i(\theta)$, the gapless directions. Second, the kinetic
quadratic term $f_0\,\tfrac{i}{2}[\phi_i,\dot\phi_i]_\star$ is
smooth in $f_0$, and one integration by parts splits it into a
contour part, carrying the jump of $f_0$, and a residual curl
density $\tfrac12\,\epsilon_{lm}\partial_l\partial_m\phi_i\,
\dot\phi_i$, which vanishes for smooth $\phi_i$ and survives only
on the singular winding configurations treated in
Sec.~\ref{sec:zeromode}. Assembling the pieces, and reducing the
energy term by cyclicity of the star product under the integral,
$\int\big[\phi_i,[\phi_i,f_0]_\star\big]_\star\,\varepsilon
=-\int[\phi_i,f_0]_\star\,[\phi_i,\varepsilon]_\star$, the action
splits into two pieces \cite{YeWang2025},
$S=\sum_{i=1}^{N_\Phi}(S_{\rm cb}+S_{\rm w})$, with
\begin{equation}
S_{\rm cb}=\int\frac{dt\,d\theta}{4\pi}\,
\partial_\theta\phi_i\Big(\dot\phi_i
-\omega(\theta)\,\partial_\theta\phi_i\Big),
\label{eq:Scb}
\end{equation}
\begin{equation}
S_{\rm w}=\int\frac{dt\,d^{2}k}{2\pi}\,f_0(\mathbf k)
\Big[-\frac{1+B\Omega}{B}\,\dot\phi_i
+\frac{1}{2}\,\epsilon_{lm}\,\partial_l\partial_m\phi_i\,
\dot\phi_i\Big].
\label{eq:Sw}
\end{equation}
The first term of Eq.~\eqref{eq:Scb} is the contour part of the
kinetic quadratic term. The second collects the reduced energy
term, both single commutators evaluated by
Eq.~\eqref{eq:new-localize}, with $\omega(\theta)$ the local drift
frequency of the field-generated band along the contour,
\begin{equation}
\omega(\theta)=\frac{B\,|\partial_{\mathbf k}\varepsilon|}
{(1+B\Omega)\,|\partial_\theta\mathbf k|}\bigg|_{\varepsilon=\mu}.
\label{eq:omtheta}
\end{equation}
Equation~\eqref{eq:Scb} is a chiral boson propagating along the
contour \cite{FJ1987}. Equation~\eqref{eq:Sw} collects the smooth
$f_0$-weighted terms: the zeroth-order time derivative and the
curl density, written with explicit indices. The term
\eqref{eq:Sw}, integrated over the occupied region, does not
affect the equations of motion. It is nonzero only on winding
configurations, and it is what carries the oscillation physics.

The boson is compact. Its mode expansion on the contour,
\begin{equation}
\phi_i(\theta)=q_i+w_i\,\theta
+\sum_{n\neq0}\frac{a_{i,n}}{\sqrt{|n|}}\,e^{in\theta},
\qquad q_i\simeq q_i+2\pi,
\label{eq:modes}
\end{equation}
contains a compact zero mode $q_i$ and an integer winding number
$w_i$. The winding is the physical charge added to the contour at
magnetic state $i$. Expanding the conjugation to first order in
$\phi_i$ and applying Eq.~\eqref{eq:new-localize}, the change in
occupation is
\begin{align}
\delta n_i&=\int\!\frac{d^{2}k}{2\pi B}\,(1+B\Omega)
\big(e^{i\phi_i}\!\star f_0\star e^{-i\phi_i}-f_0\big)
\nonumber\\
&=\int\!\frac{d^{2}k}{2\pi B}\,(1+B\Omega)\,
i\,[\phi_i,f_0]_\star+O(\phi_i^{2})
\nonumber\\
&=-\!\int_{-\pi}^{\pi}\!\frac{d\theta}{2\pi}\,\partial_\theta\phi_i
=-w_i ,
\label{eq:dn}
\end{align}
where $\theta$ increases counterclockwise, the orientation used
throughout, and the higher orders in $\phi_i$ integrate to zero on
a winding configuration. The signs are written for an electron-like
contour, and on a hole-like contour both this sign and that of the
winding term flip, and the two flips cancel in the zero-mode action
derived next. Through its winding sector the particle number can change, as it
must at fixed $\mu$.

\subsection{The zero-mode sector}
\label{sec:zeromode}

Inserting the mode expansion \eqref{eq:modes} into
$S_{\rm cb}+S_{\rm w}$ separates the action into a Fock sector of
the oscillators $a_{i,n}$ and a zero-mode sector of the pairs
$(q_i,w_i)$. For the Fermi gas and the Fermi liquid,
Ref.~\cite{Wang2025} showed that the de Haas--van Alphen
oscillations are carried by the zero-mode sector alone. The same
division holds here, and for the same reason. The Fock oscillators
$a_{i,n}$ enter the action \eqref{eq:Scb} with no coupling to the occupied-state count $n(\mu,B)$ introduced
below, so their Gaussian integral is a
prefactor of the partition function, smooth in $B$. The
$B$-periodic structure enters only through the winding numbers, the
discrete charge transfers of Eq.~\eqref{eq:dn}, which shift the
occupation against the count. We therefore keep the pairs
$(q_i,w_i)$ and derive their action.

Evaluated at $\partial_\theta\phi_i=w_i$ and
$\dot\phi_i=\dot q_i$, the first term of
Eq.~\eqref{eq:Scb} gives $\tfrac12\,w_i\dot q_i$. The
second term of Eq.~\eqref{eq:Sw} is evaluated on the winding
configuration extended into the occupied region, a field whose
angular winding is $w_i$ around an interior point; for any such
extension
\begin{equation}
\int_{\varepsilon<\mu}\!\frac{d^{2}k}{2\pi}\,
\frac{1}{2}\,\epsilon_{lm}\,\partial_l\partial_m\phi_i\,\dot q_i
=\frac{1}{2}\,w_i\,\dot q_i ,
\label{eq:new-winding}
\end{equation}
since $\epsilon_{lm}\partial_l\partial_m\phi_i$ integrates, by
Stokes' theorem, to the circulation
$\oint\partial_\theta\phi_i\,d\theta=2\pi w_i$ of the boundary
value, independent of the interior extension. Together the two
halves give the kinetic term $w_i\dot q_i$. The first term of Eq.~\eqref{eq:Sw} contributes
$-n(\mu,B)\,\dot q_i$, with
\begin{equation}
n(\mu,B)=\frac{1}{2\pi B}\!\int_{\varepsilon<\mu}\!\!(1+B\Omega)\,d^{2}k,
\label{eq:count}
\end{equation}
the occupied phase-space volume per magnetic state in the measure of
Eq.~\eqref{eq:Xidef}. The velocity term of Eq.~\eqref{eq:Scb} is
quadratic in $\partial_\theta\phi_i$. It is also the only term
that depends on how the contour is parametrized, since the kinetic term
$\partial_\theta\phi_i\,\dot\phi_i$ is reparametrization invariant (the
Jacobian of $d\theta$ cancels that of $\partial_\theta$) and the
winding numbers are integrals of $\partial_\theta\phi_i\,d\theta$.
Relabeling $\theta$ so that $\omega$ becomes uniform therefore changes
nothing else. The new parameter still spans $2\pi$, so a uniform
frequency has circuit time $2\pi/\eeff$, and equating this to the
invariant circuit time $\oint d\theta/\omega(\theta)$ gives
$\eeff=2\pi\big[\oint d\theta/\omega(\theta)\big]^{-1}$. Along the
drift $\dot\theta=\omega(\theta)$, so $\oint d\theta/\omega(\theta)$ is
the drift period $T$ around the contour and $\eeff=2\pi\hbar/T$ is the
level spacing of an orbit of period $T$. In the uniform
parametrization the winding mode has $\partial_\theta\phi_i=w_i$,
and the velocity term contributes $-\tfrac{\eeff}{2}\,w_i^{2}$. The zero-mode action is
therefore
\begin{equation}
S_{\rm zero}=\sum_{i=1}^{N_\Phi}\int dt\,
\Big[w_i\,\dot q_i\;-\;n(\mu,B)\,\dot q_i\;-\;\frac{\eeff}{2}\,w_i^{2}\Big],
\label{eq:Szero}
\end{equation}
and its two coefficients, the count $n(\mu,B)$ and the drift
frequency $\eeff$, remain to be evaluated.

The count separates into area and Berry flux,
\begin{equation}
n(\mu,B)=\frac{\Aocc(B)}{2\pi B}
+\frac{1}{2\pi}\int_{\varepsilon<\mu}\!\Omega\;d^{2}k,
\label{eq:areaflux}
\end{equation}
where the flux integral runs over the occupied region
$\{\mathbf k:\varepsilon(\mathbf k;B)<\mu\}$ and $\Aocc(B)$ is its
area, field-dependent through the second-order dispersion. With $\mu=\varepsilon_0$, the Fermi contour sits at
$\morb=-\tfrac{B}{2}\cchi|_{\mathcal C_0}+\cdots$. Each point of
$\mathcal C_0$ is displaced along $\nabla\morb/|\nabla\morb|$ by
the distance $-\tfrac{B}{2}\,\cchi/|\nabla\morb|$. Each boundary element transfers the strip of width
$\tfrac{B}{2}\,\cchi/|\nabla\morb|$ out of the occupied region where
$\cchi>0$ and into it where $\cchi<0$, for either contour family, so
\begin{equation}
\Aocc(B)=\Aocc(0)-\frac{B}{2}
\oint_{\mathcal C_0}\frac{\cchi}{|\nabla\morb|}\,d\ell.
\label{eq:afs}
\end{equation}
The zero-field value is fixed by the area $\AM$ enclosed by
$\mathcal C_0$. One has $\Aocc(0)=\AM$ for an electron-like contour, whose enclosed
$\morb<0$ side is occupied, and $\Aocc(0)=4\pi^{2}-\AM$ for a
hole-like one, whose enclosed pocket
$P=\{\mathbf k:\morb(\mathbf k)>0\}$ is unoccupied. In the
hole-like case the occupied-region flux is
$2\pi C-\int_{P}\Omega\;d^{2}k+O(B)$, with $C$ the Chern number of
the band, and the count becomes
\begin{align}
n(\mu,B)=\Big(\frac{2\pi}{B}+C\Big)
&-\frac{4\pi^{2}-\Aocc(B)}{2\pi B}
\nonumber\\
&-\frac{1}{2\pi}\int_{P}\Omega\;d^{2}k
+O(B),
\label{eq:holecount}
\end{align}
where $2\pi/B$ is the number of flux quanta per Brillouin zone. Every quantitative statement below is made at the fluxes
$B=2\pi/q$ with integer $q$, where $2\pi/B+C$ is an integer and
drops from every harmonic.
The area, the Berry flux, and the $\cchi$ correction enter the
oscillations only through the single function $n(\mu,B)$.

The frequency follows by substituting Eq.~\eqref{eq:omtheta} into
the period; with $d\ell=|\partial_\theta\mathbf k|\,d\theta$ as the
line element of the contour,
\begin{equation}
\frac{2\pi}{\eeff}
=\oint\frac{d\theta}{\omega(\theta)}
=\oint_{\varepsilon=\mu}\frac{(1+B\Omega)\,d\ell}{B\,|\partial_{\mathbf k}\varepsilon|}\,.
\label{eq:period}
\end{equation}
On the contour the gradient is
set by the moment alone,
$\partial_{\mathbf k}\varepsilon=B\,\nabla\morb+O(B^{2})$ by
Eq.~\eqref{eq:compact}, and the measure factor enters at the same
order, so to leading order
\begin{equation}
\eeff \;=\;\frac{2\pi B^{2}}{\oint_{\mathcal C_0}d\ell/|\nabla\morb|}\,,
\label{eq:heff}
\end{equation}
quadratic in $B$. One power is explicit in Eq.~\eqref{eq:omtheta},
and $|\partial_{\mathbf k}\varepsilon|$ carries the second, the
dispersion being itself of first order. The $B^{2}$ scale of the
spacing was identified in Ref.~\cite{HwangRhimYang2021} by applying
the Onsager rule to the effective dispersion $B\,\morb(\mathbf k)$,
whose effective mass grows as $1/B$. Equation~\eqref{eq:heff} supplies
the coefficient in closed form. The period integral in
Eq.~\eqref{eq:period} is exactly the $\mu$ derivative of the count.
Raising $\mu$ in Eq.~\eqref{eq:count} advances the contour locally by
$d\mu/|\partial_{\mathbf k}\varepsilon|$, each element weighted by the
measure $1+B\Omega$, so that
$\oint_{\varepsilon=\mu}(1+B\Omega)\,d\ell/(B|\partial_{\mathbf k}\varepsilon|)
=2\pi\,dn/d\mu$ and $\eeff=(dn/d\mu)^{-1}$ to all orders in $B$, the
inverse density of states per magnetic state. This is the expected
relation between level spacing and density of states.

With both coefficients evaluated, the spectrum follows. For each
magnetic state, Eq.~\eqref{eq:Szero} is the quantum mechanics of a
particle on a ring threaded by a flux
\cite{YeWang2025,AltlandSimons2010}. Here $q_i$ is the compact
coordinate, the count $n(\mu,B)$ plays the role of the flux, and
$\eeff$ plays the role of the charging energy. The coefficient of $\dot q_i$ in
Eq.~\eqref{eq:Szero} identifies the momentum conjugate to $q_i$ as
$w_i-n(\mu,B)$, and compactness of $q_i$, of period $2\pi$,
quantizes this momentum to integer eigenvalues $l_i$. On such an
eigenstate $w_i=l_i+n(\mu,B)$ (integer windings label configurations of the path
integral, not eigenstates), and the quadratic term of
Eq.~\eqref{eq:Szero} is
\begin{equation}
\frac{\eeff}{2}\,w_i^{2}
=\frac{\eeff}{2}\,\big(l_i+n(\mu,B)\big)^{2}.
\end{equation}
The partition function per magnetic state is
\begin{equation}
Z_{\rm zero}=\sum_{l\in\mathbb Z}
\exp\!\Big[-\frac{\eeff}{2T}\,\big(l+n(\mu,B)\big)^{2}\Big],
\label{eq:Zzero}
\end{equation}
periodic in the count with unit period. The discreteness of the
spectrum enters through the integer eigenvalues of the conjugate
momentum, without a Landau-level sum.

\subsection{Free energy of the zero mode: LK series and the quantization phase}
\label{sec:LKphase}

The oscillatory grand potential is the free energy of the zero-mode
sector, $\delta\Xi_{\rm osc}=-(N_\Phi/L^{2})\,T\ln Z_{\rm zero}$, with
$N_\Phi=BL^{2}/2\pi$ the number of magnetic states. The
$B$-periodic structure is extracted by Poisson resummation of the
ladder \eqref{eq:Zzero},
\begin{equation}
Z_{\rm zero}=\sqrt{\frac{2\pi T}{\eeff}}\,
\sum_{p\in\mathbb Z}
e^{-2\pi^{2}p^{2}T/\eeff}\;e^{2\pi i p\,n(\mu,B)},
\label{eq:poisson}
\end{equation}
which trades the winding sum for a sum over harmonics. Define
\begin{equation}
X_p=\frac{2\pi^{2}pT}{\eeff},
\label{eq:new-Xp}
\end{equation}
so the Gaussian weight of harmonic $p$ in Eq.~\eqref{eq:poisson} is
$e^{-pX_p}=e^{-p^{2}X_1}$. The $p$-sum of Eq.~\eqref{eq:poisson} is a Jacobi theta function,
\begin{align}
Z_{\rm zero}&=\sqrt{\frac{2\pi T}{\eeff}}\;
\theta_3\big(\pi\,n(\mu,B),\,e^{-X_1}\big),
\nonumber\\
\theta_3(\zeta,e^{-X_1})&=\sum_{p\in\mathbb Z}
e^{-X_1p^{2}}e^{2ip\zeta},
\label{eq:new-theta}
\end{align}
whose triple product
\cite{WhittakerWatson1927,YeWang2025} is
\begin{equation}
\theta_3(\zeta,e^{-X_1})=\prod_{m\ge1}
\big(1-e^{-2mX_1}\big)\,
\big|1+e^{-(2m-1)X_1}e^{2i\zeta}\big|^{2},
\label{eq:tripleproduct}
\end{equation}
with the first factor independent of $\zeta$. Expanding the
logarithm of the second factor at $\zeta=\pi\,n(\mu,B)$,
$2\,\mathrm{Re}\ln(1+y)$ with
$\ln(1+y)=-\sum_{p\ge1}(-y)^{p}/p$, the real part supplies
$\cos(2\pi p\,n(\mu,B))$. At each $p$ the sum over $m$ is geometric,
$\sum_{m\ge1}e^{-(2m-1)X_p}=(2\sinh X_p)^{-1}$, and
\begin{equation}
-T\ln Z_{\rm zero}
=T\sum_{p\ge1}\frac{(-1)^{p}}{p\sinh X_p}
\cos\!\big(2\pi p\,n(\mu,B)\big),
\label{eq:harmonic}
\end{equation}
up to smooth terms, the Gaussian prefactor of
Eq.~\eqref{eq:poisson} and the first factor of
Eq.~\eqref{eq:tripleproduct}, which join the nonoscillatory
thermodynamics.
The alternating sign is absorbed into the phase,
$(-1)^{p}\cos(2\pi p\,n)=\cos\!\big[2\pi p\,(n+\tfrac12)\big]$. The
half-integer offset of the ladder is the origin of the Maslov
one-half, here a property of the dual series rather than a
quantization input. By
Eq.~\eqref{eq:holecount} the integer $2\pi/B+C$ drops from every
harmonic, the cosine is even, and by Eq.~\eqref{eq:afs} the pocket
term of Eq.~\eqref{eq:holecount} splits into the $1/B$ frequency
term $\AM/2\pi B$ and an $O(1)$ shift. The free energy assembles into
\begin{equation}
\delta\Xi_{\mathrm{osc}}
=\frac{B\,\eeff}{4\pi^{3}}
\sum_{p\ge1}\frac{X_p}{p^{2}\sinh X_p}
\cos\!\Big[2\pi p\Big(\frac{\AM}{2\pi B}-\tilde\gamma\Big)\Big],
\label{eq:LK}
\end{equation}
the LK form, with the oscillation frequency $\AM/2\pi$ read
off Eq.~\eqref{eq:holecount} as the pocket area over $2\pi$, and
the quantization phase collecting the three $O(1)$ pieces of
the count,
\begin{equation}
\tilde\gamma \;=\; \underbrace{\frac12}_{\text{Maslov}}
\;\underbrace{-\,\frac{1}{2\pi}\int_{P}\Omega\;d^{2}k}_{\tilde\gamma_1
\text{ (Berry)}}
\;\underbrace{-\,\frac{1}{4\pi}\oint_{\mathcal C_0} \frac{\cchi}{|\nabla\morb|}\,d\ell}_{\tilde\gamma_2\ (\cchi\text{ term})} .
\label{eq:gamma}
\end{equation}
Equations~\eqref{eq:LK}--\eqref{eq:gamma} are the central result
of this paper: a closed Lifshitz--Kosevich formula for a band with
no zero-field Fermi surface, every entry built from the zero-field
wavefunctions. Each term has its own origin. The one-half is the half-integer offset of
the dual series. The Berry term $\tilde\gamma_1$ comes from the
measure, the Berry flux accumulated over the pocket interior. The $\cchi$ term
$\tilde\gamma_2$ is the first-order shift of the quantized contour
produced by the $O(B^{2})$ dispersion, surviving as an $O(1)$ phase
precisely because of the quadratic spacing \eqref{eq:heff}. Its
analog in an ordinary metal is the Roth correction \cite{Roth1966},
small and usually neglected. The prefactor $B\,\eeff/4\pi^{3}$ falls as $B^{3}$, one power
from the level degeneracy $B/2\pi$ and two from the spacing
$\eeff$. A metallic spacing, linear in $B$, gives $B^{2}$.
Equation~\eqref{eq:gamma} is written for the hole-like pocket used
above. For an electron-like contour the count needs no zone term.
It is Eq.~\eqref{eq:areaflux} as it stands, with the flux integral
over the occupied pocket $P$. The harmonics of
Eq.~\eqref{eq:harmonic} are even in $n$, so the hole-like series
may equally be generated by $-n$ of Eq.~\eqref{eq:holecount}.
Comparing $-n$ with Eq.~\eqref{eq:areaflux} term by term,
the leading area term and the Berry term agree, the integer drops
from the harmonics as before, and the shift of Eq.~\eqref{eq:afs}
enters once directly and once through the complement
$4\pi^{2}-\Aocc(B)$, with opposite sign.
Every step from Eq.~\eqref{eq:harmonic} to Eq.~\eqref{eq:gamma}
then repeats verbatim, and only the $\cchi$ term of the phase
changes sign:
\begin{equation}
\tilde\gamma_2 \;=\; \pm\,\frac{1}{4\pi}
\oint_{\mathcal C}\frac{\cchi}{|\nabla\morb|}\,d\ell ,
\label{eq:signrule}
\end{equation}
the upper sign for an electron-like contour and the lower, the case
of Eq.~\eqref{eq:gamma}, for a hole-like one. In both cases $P$ in
$\tilde\gamma_1$ is the region the contour bounds (occupied for an
electron-like contour, unoccupied for a hole-like one). The hole-like case is the one tested numerically in
Sec.~\ref{sec:dhva-num}.

\section{Numerical checks}
\label{sec:numerics}

The analytic results above are parameter-free. Every predicted number
is computed from zero-field wavefunctions. This section checks them
against exact diagonalization (ED) of a lattice model.

\subsection{Lattice model and numerical prescription}
\label{sec:model-num}

\begin{figure}[tb]
\includegraphics[width=0.74\columnwidth]{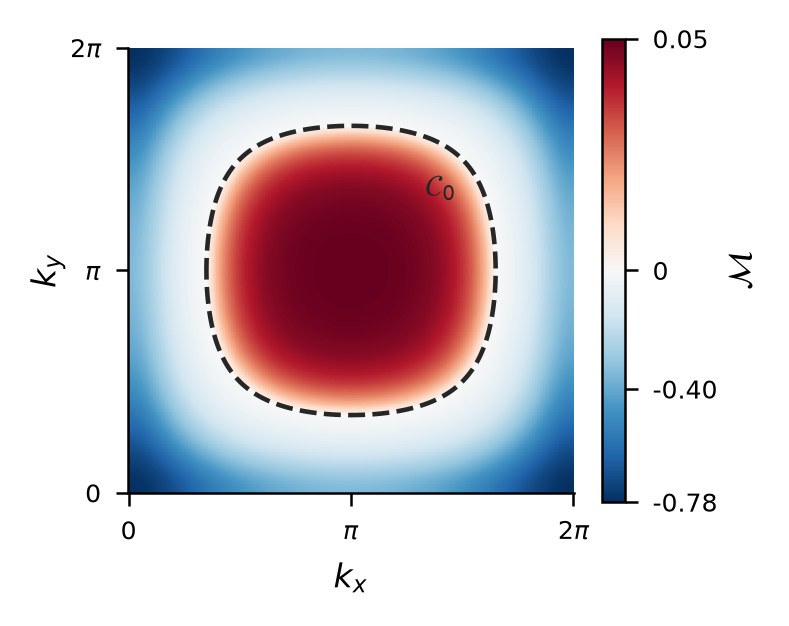}
\caption{The orbital moment $\morb(\mathbf k)$ of the lower band at
$u=-1.2$, in the zone chart centered on $M=(\pi,\pi)$. The moment
is negative at $\Gamma$ and $X$, with minimum $-0.78$ at $\Gamma$,
and positive inside a single pocket around $M$, with maximum
$+0.05$. The color scale is linear on each side of zero. Dashed:
the zero contour $\mathcal C_0$, the flat-band Fermi surface of
Sec.~\ref{sec:LK}.}
\label{fig:moment}
\end{figure}

We take the flattened two-band model
\begin{equation}
\widehat H(\mathbf k)=\hat d(\mathbf k)\cdot\bm\sigma,
\qquad
\hat d=\frac{\mathbf d}{|\mathbf d|},
\label{eq:model}
\end{equation}
with $\mathbf d=(\sin k_x,\ \sin k_y,\ u+\cos k_x+\cos k_y)$ the
Qi--Wu--Zhang vector \cite{QWZ2006}. Normalizing $\mathbf d$ flattens
both bands exactly: $\varepsilon_\pm=\pm1$, full gap $\Delta=2$, leaving all $\mathbf k$-dependence in the wavefunctions. We study the
lower band at $u=-1.2$. The band is isolated for $u\neq0,\pm2$. The two bands are related by
$\varepsilon\to-\varepsilon$, so the lower band has
$\morb(\mathbf k)=\Omega(\mathbf k)$ exactly, a convenience of the two-band limit, not an input to the theory. Figure~\ref{fig:moment} shows the moment over the zone; it is negative
at $\Gamma$ and $X$, positive inside a single closed pocket around
$M$ bounded by the zero contour $\mathcal C_0$. The pocket area
$\AM$, the area enclosed by $\mathcal C_0$, and the contour integral
$\oint_{\mathcal C_0}d\ell/|\nabla\morb|$, which set the frequency
and the level spacing of the LK formula respectively, are evaluated by contour
integration on dense momentum grids, extrapolated to zero grid
spacing.

The two-band corollary \eqref{eq:chi2band} applies to this model
and is used repeatedly below. The ED
prescription used for all figures: a rational flux
$B=2\pi/q$ enters through straight-line Peierls substitution on the
real-space hoppings of the flattened model, truncated at a hopping
range where the exponential tails fall below
$10^{-12}$ per element ($|\mathbf r|\le 30$ at $u=-1.2$). The resulting $2q\times2q$
Harper matrices are diagonalized exactly for $q$ up to $2560$, with
Hermiticity residuals below $10^{-11}$. The $q$ levels that evolve
from the flat band, separated from the rest of the spectrum by the
gap, are the flat-band Landau cluster of the checks below.

\subsection{The field-generated spectrum}
\label{sec:band-num}

\begin{figure}[tb]
\includegraphics[width=\columnwidth]{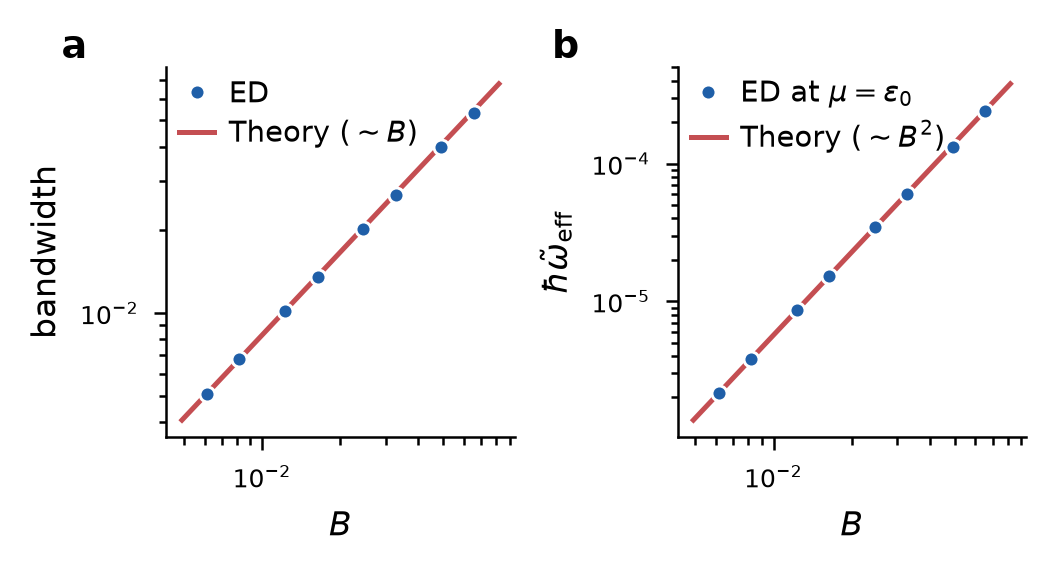}
\caption{Log-log plots for the field-generated band at $u=-1.2$. (a) ED bandwidth of
the flat-band Landau cluster vs $B$ (points) against
$(\morb^{\max}-\morb^{\min})B$ (line, no fit). (b) ED level spacing
at $\mu=\varepsilon_0$ vs $B$ (points) against
$\eeff=2\pi B^{2}\big/\!\oint_{\mathcal C_0}d\ell/|\nabla\morb|$
[Eq.~\eqref{eq:heff}; line, no fit].}
\label{fig:band}
\end{figure}

Figure~\ref{fig:band} shows the two structural scalings. The ED
bandwidth follows $(\morb^{\max}-\morb^{\min})B$, and the level
spacing at $\mu=\varepsilon_0$ follows the quadratic law of
Eq.~\eqref{eq:heff} with the independently computed contour integral
$\oint_{\mathcal C_0}d\ell/|\nabla\morb|=110.65$, both without
adjustable parameters.
The spacing law involves only $\oint d\ell/|\nabla\morb|$ and is therefore
independent of $\cchi$.

\subsection{The smooth response}
\label{sec:chi-num}

\begin{figure}[tb]
\includegraphics[width=0.9\columnwidth]{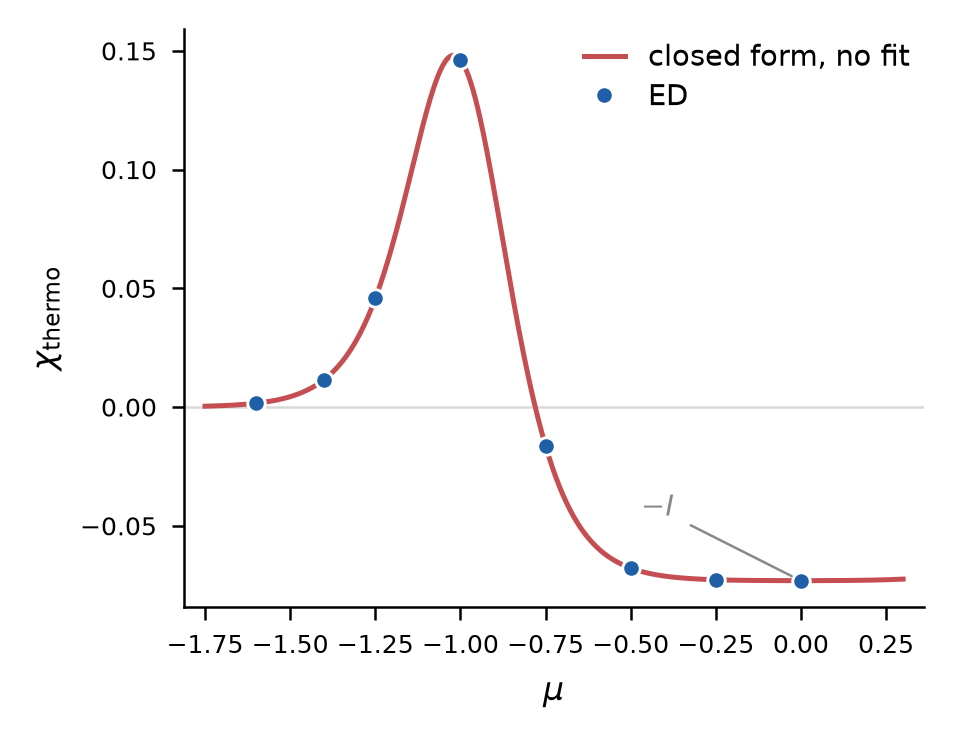}
\caption{Orbital magnetic susceptibility of the flat-band model at
$u=-1.2$, $T=0.1$: ED (points; quadratic coefficient of the smooth
grand potential, $q=320$--$1200$) against the closed form
Eq.~\eqref{eq:chiclosedform} (line, no fit). The paramagnetic peak at
$\mu=\varepsilon_0$ is the orbital Curie response of the
field-generated moments. The plateau in the gap is the geometric diamagnetism
$\chi_{\rm thermo}=-I$, with no Fermi surface.}
\label{fig:chi}
\end{figure}

The smooth part of the ED grand potential is fitted by a quartic polynomial in
$B$ over $q\in[320,1200]$ at $T=0.1$, where the dHvA terms are
thermally negligible, and the linear
and quadratic coefficients are compared with $-M$
[Eq.~\eqref{eq:Mthermo}] and $-\chi_{\rm thermo}/2$
[Eq.~\eqref{eq:chiclosedform}]. Figure~\ref{fig:chi} shows the
$\mu$-scan at $T=0.1$. The ED tracks the closed form through the Curie peak at the band
and through the in-gap diamagnetic plateau
($\chi_{\rm thermo}=-I$ to $0.1\%$ at midgap, with $I=0.0732$ at
$u=-1.2$).

\subsection{The dHvA effect with $\mu$ pinned to the flat band}
\label{sec:dhva-num}

\begin{figure}[tb]
\includegraphics[width=\columnwidth]{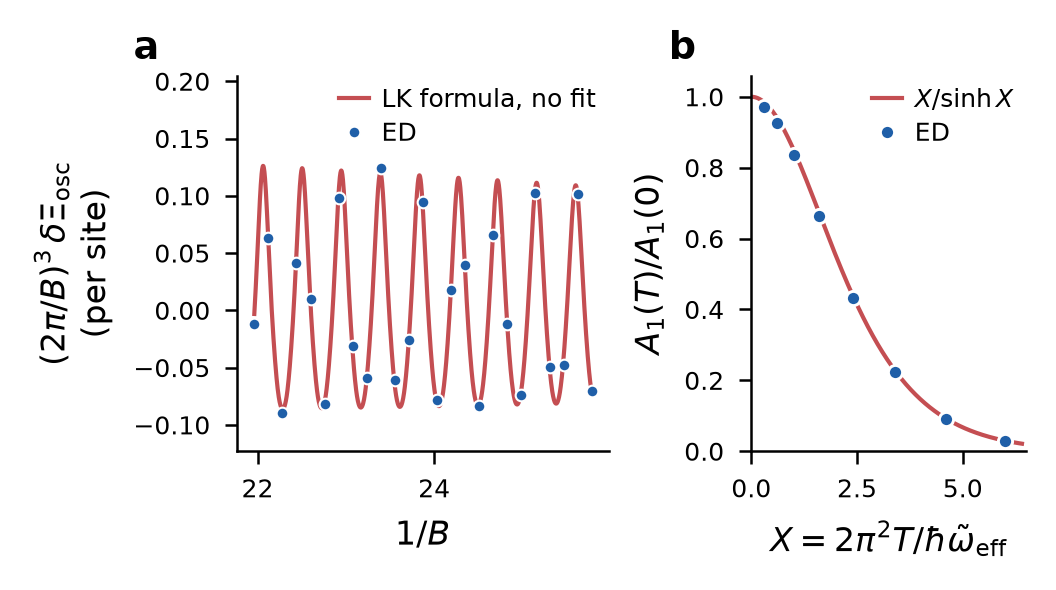}
\caption{dHvA oscillations at $u=-1.2$.
(a) Oscillatory grand potential vs $1/B$ at pinned
$\mu=\varepsilon_0$ and $T=5\times10^{-6}$: ED points after
subtracting a polynomial fit of the smooth part in $B$, rescaled by
$(2\pi/B)^{3}$ to cancel the $B^{3}$ envelope of Eq.~\eqref{eq:LK},
against the LK formula with all parameters fixed by geometry and
$\tilde\gamma=0.874$ (line, no fit). (b) Fundamental amplitude $A_1$ vs $T$ on the in-band
contour $\morb=-\tfrac12$ ($\mu=\varepsilon_0-B/2$, $q=150$),
normalized to $T\to0$, against $X/\sinh X$ with
$X=2\pi^{2}T/\eeff$ and $\eeff$ from Eq.~\eqref{eq:heff} evaluated
on that contour.}
\label{fig:dhva}
\end{figure}

Figure~\ref{fig:dhva} shows the LK series against ED; both the waveform
against $1/B$ at pinned $\mu=\varepsilon_0$, with frequency
$\AM/2\pi=2.26$, and the fundamental amplitude against
$X/\sinh X$ damping on an in-band contour, both with every
parameter fixed by geometry. The quantization phase is tested by the same waveform. The curve of
Fig.~\ref{fig:dhva}(a) uses the zero-parameter prediction
$\tilde\gamma=0.874$ of Eq.~\eqref{eq:gamma}, with pieces
$\tilde\gamma_1=-0.078$ and $\tilde\gamma_2=+0.452$. Without the
$\cchi$ term, the phase would be $0.422$, nearly half a period away.
On this pocket the $\cchi$ term is the larger, by a factor of $5.8$.

\begin{figure*}[tb]
\includegraphics[width=0.86\textwidth]{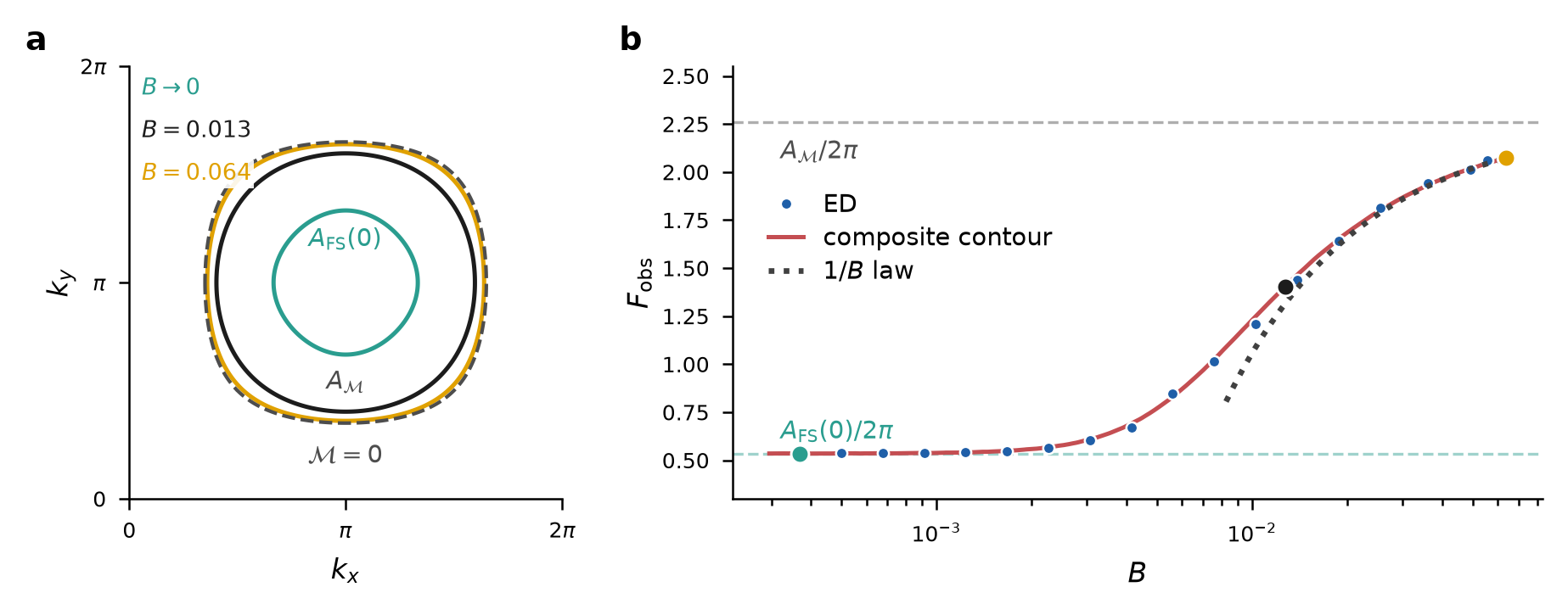}
\caption{The dHvA crossover of a narrow band added to the model of
Sec.~\ref{sec:numerics} ($u=-1.2$, $W=2.8\times10^{-4}$,
$\mu-\varepsilon_0=3.5W$, half a $W$ below the band top; the whole sweep sits inside the
expansion window, $x=B\xi^{2}\le0.10$ with $\xi^{2}=1.6$). (a)~The quantized contour
$Wg(\mathbf k)+B\,\morb(\mathbf k)=\mu-\varepsilon_0$ in the zone: at
$B\to0$ the zero-field Fermi surface, a hole pocket around $M$ with
area $A_{\rm FS}(0)$. With increasing field it inflates onto the
$\morb=0$ contour (dashed, area $\AM$), the flat-band Fermi
surface of Sec.~\ref{sec:LK}. (b)~The frequency it produces:
instantaneous frequency, the windowed slope of the level count
$N(\mu;q)$ over $25$--$75$ contiguous flux values (points; counts
by matrix inertia for $q$ up to $1.7\times10^{4}$), against the
composite-contour prediction of Eq.~\eqref{eq:Fobs} (line, no fit),
rising from the plateau at $A_{\rm FS}(0)/2\pi$ toward the asymptote
$\AM/2\pi$ (dashed) along the $1/B$ law of
Eq.~\eqref{eq:oneoverB} (dotted). Colored dots in (b) mark the
fields of the contours in (a), the chirp midpoint and the sweep
edge.}
\label{fig:crossover}
\end{figure*}

\section{A narrow band: crossover into the flat-band regime}
\label{sec:crossover}

Any flat band realized in a material retains a small dispersion,
and the oscillations of Sec.~\ref{sec:LK} apply where the
field-generated dispersion dominates it. At weak field the
oscillations are the ordinary dHvA of the narrow band. Between the
two limits lies a crossover, and its profile follows from the same
construction with no new input. Let the band carry a small
built-in dispersion,
$E_0(\mathbf k)=\varepsilon_0+W\,g(\mathbf k)$ with
$g$ of order unity and bandwidth $W\ll\Delta$. The field
expansion of Sec.~\ref{sec:dispersion} is organized by
$x=B\xi^{2}$, a wavefunction property independent of $W$, and it
is exact at this order when $W=0$, so every term coupling the
built-in dispersion to the field carries at least one power of
$W$. At the retained order the two dispersions add:
\begin{equation}
\varepsilon_W(\mathbf k;B)=Wg(\mathbf k)+B\,\morb(\mathbf k)
+\frac{1}{2}\,\cchi(\mathbf k)\,B^{2}+O(Wx,\Delta x^{3}),
\label{eq:composite}
\end{equation}
measured from $\varepsilon_0$, with $\morb$ and $\cchi$ the
flat-band functionals of Eqs.~\eqref{eq:morb}
and~\eqref{eq:chifull}. The terms by which the second-order
energy of the composite band differs from $\tfrac12\cchi B^{2}$
all carry $W$ and sit in the $O(Wx)$ error. The built-in term is field
independent while the geometric term grows linearly in $B$, and the
larger of the two changes at the crossover field
% AUTHOR DECISION (audit + annotation 'what does max_P mean?'): replace the
% occupied-region extrema max_P/min_P with zone extrema g^max-g^min over
% M^max-M^min (shifts the value 0.008 -> 0.010; kills the
% P collision with Sec IV D). Say the word and it goes in.
\begin{equation}
B_{\morb}\;=\;W\,
\frac{\max_{P}g-\min_{P}g}{\max_{P}\morb-\min_{P}\morb},
\label{eq:new-Bstar}
\end{equation}
where both ranges are taken over the occupied region at the
filling considered, at weak field the zero-field Fermi sea. Landau
quantization proceeds on the \emph{composite} contours. The
construction of Sec.~\ref{sec:LK} quantizes the level sets of the
first-order dispersion by the area rule, the second-order term
entering only through the phase $\tilde\gamma_2$ of
Eq.~\eqref{eq:signrule}. It applies verbatim with $Wg+B\morb$ in
place of $B\morb$; the same area rule and quantization phase, now
on contours of the composite first-order dispersion rather than of
$\morb$ alone.

The observable consequence is a frequency chirp. The
oscillations stay locally periodic in $1/B$ while their frequency
drifts across the sweep. At fixed $\mu=\varepsilon_0+W g_0$ the
quantized contour is the level set $Wg+B\morb=W g_0$. The pocket it bounds, here the unoccupied
region, has area $A_{\rm FS}(B)=4\pi^{2}-\Aocc(B)$, and by
Eq.~\eqref{eq:holecount} it sets the phase of every harmonic of
Eq.~\eqref{eq:LK}, the integer zone term $2\pi/B+C$ dropping as in
Sec.~\ref{sec:LKphase}. A measurement sweeping $1/B$ reads the
local frequency
\begin{equation}
F_{\rm obs}(B)\;=\;\frac{d}{d(1/B)}
\Big[\frac{A_{\rm FS}(B)}{2\pi B}\Big],
\label{eq:new-Fobsdef}
\end{equation}
and carrying out the derivative gives
\begin{equation}
F_{\rm obs}(B)\;=\;\frac{1}{2\pi}
\Big[A_{\rm FS}(B)\;-\;B\,\frac{dA_{\rm FS}}{dB}\Big],
\label{eq:Fobs}
\end{equation}
where the derivative term is the contribution of the field-dependent
area. Where the field-generated term shifts the
pocket area appreciably, the derivative term is comparable, and
the measured frequency separates from the instantaneous
$A_{\rm FS}(B)/2\pi$. The quantization phase $\tilde\gamma$ of the composite contour
also drifts with $B$. It contributes $B^{2}\,d\tilde\gamma/dB$
to the measured frequency, its accumulated phase over the sweep is
bounded by the $O(1)$ range of $\tilde\gamma$, and it is negligible
against the $1/B$ frequency terms throughout. The two limits of Eq.~\eqref{eq:Fobs} follow from
perturbing the contour about its two endpoints. For $B\ll B_{\morb}$
the contour is the zero-field Fermi surface of the narrow band, and
the pocket area drifts linearly in the field-generated term,
\begin{align}
A_{\rm FS}(B)\;&=\;A_{\rm FS}(0)\;+\;\frac{B}{W}
\oint_{\rm FS}\frac{\morb}{|\nabla g|}\,d\ell,
\nonumber\\
F_{\rm obs}(B)\;&=\;\frac{A_{\rm FS}(0)}{2\pi}\;+\;O(B^{2}),
\label{eq:new-weak}
\end{align}
the drift cancelling between the two terms of Eq.~\eqref{eq:Fobs}; this is
ordinary dHvA, with the zero-field frequency protected to linear
order. For $B\gg B_{\morb}$ the
contour condition becomes $\morb=(W/B)(g_0-g)\to0$. The
contour flows to $\mathcal C_0$, and the
oscillations become the flat-band oscillations of
Sec.~\ref{sec:LK}, frequency and phase included. The approach to the asymptote is also predicted. Perturbing the contour off
$\mathcal C_0$ gives
\begin{align}
A_{\rm FS}(B)\;&=\;\AM\;-\;\frac{W}{B}\,\Lambda_g,
\nonumber\\[2pt]
F_{\rm obs}(B)\;&=\;\frac{1}{2\pi}
\Big[\AM\;-\;\frac{2\,\Lambda_g\,W}{B}\Big],
\label{eq:oneoverB}
\end{align}
where $\Lambda_g=\oint_{\mathcal C_0}(g_0-g)\,d\ell/|\nabla\morb|$ and
$\AM$ is the area enclosed by $\mathcal C_0$. The tail's
coefficient is a zero-field contour integral, and its factor of
two relative to the area law is an artifact of the derivative term in Eq.~\eqref{eq:Fobs}. The same composite-contour
rule fixes the other observables across the crossover. The level
spacing interpolates between the linear-in-$B$ law of a dispersive band and the quadratic law \eqref{eq:heff}, and
the quantization phase interpolates between its endpoint values on
the two contour families.

The quantization phase migrates with the contour, but the phase a
measurement reads has its own scale, parametrically later than
$B_{\morb}$. Inserting Eq.~\eqref{eq:oneoverB} into the phase of
each harmonic of Eq.~\eqref{eq:LK} splits it into three terms,
\begin{align}
2\pi p&\Big[\frac{A_{\rm FS}(B)}{2\pi B}-\tilde\gamma\Big]
\nonumber\\
&=p\,\frac{\AM}{B}\;-\;p\,\frac{W\Lambda_g}{B^{2}}
\;-\;2\pi p\,\tilde\gamma,
\label{eq:new-phasesplit}
\end{align}
the frequency term, the dispersion residual, and the geometric
offset. The residual and the offset are equal at
\begin{equation}
B_{\tilde\gamma}=\Big[\frac{W\,\Lambda_g}
{2\pi\,|\tilde\gamma_1+\tilde\gamma_2|}\Big]^{1/2},
\label{eq:new-Bgamma}
\end{equation}
where $\tilde\gamma_{1,2}$ are their $\mathcal C_0$ values. The
Maslov half, common to every dHvA series, drops from the
comparison. Both terms scale as $p$, so $B_{\tilde\gamma}$ is the same for
every harmonic. Beyond $B_{\tilde\gamma}$, the residual dispersion averaged
over $\mathcal C_0$ is smaller than the geometric fraction
$|\tilde\gamma_1+\tilde\gamma_2|$ of the level spacing $\eeff$, and
the phase is carried by its Berry and $\cchi$ terms. The two scales balance the built-in dispersion against the two
orders of the field expansion, the bandwidth of the first-order
term at $B_{\morb}$ and the level spacing $\eeff$ at
$B_{\tilde\gamma}$, so $B_{\morb}\propto W$ while
$B_{\tilde\gamma}\propto W^{1/2}$. The window
$B_{\morb}\ll B\ll B_{\tilde\gamma}$, with the frequency saturated
but the phase still dispersion-dominated, widens as the band
flattens.

We verify the crossover by exact diagonalization on the model of
Sec.~\ref{sec:numerics} augmented by a scalar dispersion
$W\,g(\mathbf k)\,\sigma^{0}$, with $\sigma^{0}$ the identity
in orbital space and $g=2-\cos k_x-\cos k_y$,
chosen because it commutes with the geometric part, so $\morb$ and
$\cchi$ are exactly unchanged while the band acquires the bandwidth
$4W$. The chemical potential is placed just below the top of the
narrow band, and the bandwidth is chosen so that the whole field sweep
stays inside the expansion window $x=B\xi^{2}\ll1$ (parameters in the
caption of Fig.~\ref{fig:crossover}). The zero-field Fermi surface is a hole pocket
around $M$, and with growing field it inflates onto the same
$\mathcal C_0$ whose oscillations Secs.~\ref{sec:LK} and
\ref{sec:numerics} quantify, the quantized contour staying hole-like
for the whole sweep. Figure~\ref{fig:crossover} shows the measured
instantaneous frequency across the sweep. It follows the
composite-contour prediction of Eq.~\eqref{eq:Fobs} with no fitted
parameters, from the low-field plateau at $A_{\rm FS}(0)/2\pi$ through
the full chirp. The plateau shows the cancellation of
Eq.~\eqref{eq:new-weak}. Deep in the plateau the pocket area has
already grown by several tens of percent while the measured frequency
has barely moved. The $1/B$ law of
Eq.~\eqref{eq:oneoverB} carries the strong-field
end, where the measured frequency still sits visibly below the
asymptote, by the residual $2\Lambda_gW/B$ at the edge of the
sweep. The same sweeps test the
spacing interpolation. The measured level-spacing exponent crosses
over from $1$ in the weak-field window to $2$ in the strong-field
window, as the composite contour predicts.

The phase scale $B_{\tilde\gamma}$ of
Eq.~\eqref{eq:new-Bgamma} lies a factor of two beyond the
strong-field edge of the sweep for this protocol, so the
frequency sweep does not test the phase. The phase is measured
separately, per flux value, from the level positions, as the
fractional index of the level nearest $\mu$ within the Landau
cluster, referred to the composite pocket area $A_{\rm FS}(B)$ and
differenced against a $W=0$ sweep on the same flux grid. The measured phase follows
Eqs.~\eqref{eq:gamma} and~\eqref{eq:signrule} evaluated on the
composite contour at every field, migrating from the narrow-band
value to the $\mathcal C_0$ value of Sec.~\ref{sec:dhva-num}, more
than a third of a period away, with the midpoint of the migration
inside the chirp, and it has essentially reached the $\mathcal C_0$
value at $B_{\tilde\gamma}$. What changes there is the composition of a
phase reading. A constant-frequency fit carries the dispersion
residual $W\Lambda_g/2\pi B^{2}$ of Eq.~\eqref{eq:new-phasesplit}
in its intercept, alongside the geometric offset, and the measured
contamination crosses the geometric content at $B_{\tilde\gamma}$,
as Eq.~\eqref{eq:new-Bgamma} prescribes.

\section{Conclusions}
\label{sec:conclusions}

In this paper, we have shown that a band with no dispersion has an
orbital response determined entirely by wavefunction geometry. The magnetic field generates the
dispersion, the moment term $B\,\morb(\mathbf k)$ at first order and
the coefficient $\cchi(\mathbf k)$ at second, and the single
parameter $x=B\xi^{2}$ controls the expansion. The susceptibility that follows is purely geometric and consistent
with the exact resolvent response formula for any isolated,
nondegenerate flat band. Bosonizing
the emergent zero contour gives a modified Lifshitz--Kosevich
formula, obtained without invoking the semiclassical quantization
rule of cyclotron orbits, whose quantization phase carries the
second-order geometry at $O(1)$. In a narrow band the dHvA frequency
and the quantization phase migrate from their zero-field values to
the flat-band ones at the crossover field $B_{\morb}\propto W$. The
phase content of the signal becomes geometric only at the
parametrically later $B_{\tilde\gamma}\propto W^{1/2}$, and below
that scale the offset a dHvA fit returns is dominated by the
dispersion residual rather than by the Berry and $\cchi$ terms.

Experimentally, the quadratic spacing law sets the observation
window. Oscillations require
$T\lesssim\eeff/2\pi^{2}
=B^{2}\big/\pi\!\oint_{\mathcal C_0}d\ell/|\nabla\morb|$. For a flat
band with $\xi$ a few lattice constants
($\oint d\ell/|\nabla\morb|\sim10^{2}$ in lattice units, as for the
model here), a laboratory field of a flux quantum per hundred unit
cells ($B\sim0.06$), and a typical gap of tens of meV as the unit of
energy, the window is millikelvin temperatures, standard for dHvA
experiments. Disorder is expected to enter as a Dingle factor on the same
harmonics \cite{Shoenberg1984,Wang2025}, to be compared with
$\eeff$, the more stringent constraint for flat bands. Of the observables, the magnetization
phase is the most demanding. The most accessible geometric channel
is the smooth susceptibility plateau, which requires only $T\ll\Delta$.

Flat-band platforms with tunable geometry, moir\'e systems foremost,
are the natural place to look. Interactions, which flat bands
amplify, are beyond the present noninteracting treatment. For the
smooth response a concrete interacting bridge exists through
mean-field bands \cite{KWV2025}, while an interacting theory of the
oscillatory response remains open.

\appendix

\section{Consistency with the gauge-invariant susceptibility}
\label{app:consistency}

This appendix reduces the exact trace formula of
Ref.~\cite{Raoux2015}, restated as Eq.~\eqref{eq:master} (the
noninteracting limit of the interacting formulation of
Ref.~\cite{KWV2025}), at an arbitrary isolated, nondegenerate flat
band of an $N$-band Hamiltonian, and shows that it reproduces the
structure and the coefficients of Eq.~\eqref{eq:chithermo}.

Split the resolvent at the flat energy
($P=|v^{0}\rangle\langle v^{0}|$ the flat-band projector, $Q=1-P$):
\begin{equation}
G(E)=\frac{P}{E-\varepsilon_0+i0}+R(E),\qquad
R(E)\equiv Q\,(E-h)^{-1}Q,
\label{eq:gsplit}
\end{equation}
with $R(E)$ analytic inside the flat band's spectral window, the
interval of half-width $\Delta$ around $\varepsilon_0$ that contains
no other spectrum. We write $R_0\equiv R(\varepsilon_0)$.

Differentiating the projector form of the eigenvalue equation,
$hP=\varepsilon_0P$, gives
$h^{a}P+h\,\partial_aP=\varepsilon_0\,\partial_aP$. Multiply by
$P$ from both sides:
\begin{equation}
P\,h^{a}P+P\,h\,\partial_aP\,P
=\varepsilon_0\,P\,\partial_aP\,P .
\label{eq:velid0}
\end{equation}
Since $Ph=\varepsilon_0P$, the second term on the left equals the
right-hand side, and the two cancel,
\begin{equation}
P\,h^{a}P=0 .
\label{eq:velid}
\end{equation}
A flat band carries no velocity matrix element.

Now insert Eq.~\eqref{eq:gsplit} into the trace of
Eq.~\eqref{eq:master}. Each of the four resolvent slots contributes
either the pole term $P/(E-\varepsilon_0+i0)$ or the analytic term
$R(E)$, so the trace organizes by powers of the pole factor,
\begin{equation}
\operatorname{Tr}\{\cdots\}
=\frac{\kappa_2(\mathbf k)}{(E-\varepsilon_0+i0)^{2}}
+\frac{\kappa_1(\mathbf k)}{E-\varepsilon_0+i0}
+(\mathrm{analytic}),
\label{eq:polestruct}
\end{equation}
and this display defines $\kappa_{1,2}$. Here $\kappa_2$ collects
the placements with two $P$'s and the remaining slots at $R_0$.
The coefficient $\kappa_1$ collects the placements with one $P$, together with the
two-$P$ placements in which one $R(E)$ is replaced by its first
Taylor coefficient at $\varepsilon_0$ (the factor
$E-\varepsilon_0$ so produced reduces the double pole to a simple
one). Higher poles are absent. A pole of order $p$ requires $p$
factors of $P$, any two $P$'s in cyclically adjacent slots sandwich
a velocity vertex, which vanishes by Eq.~\eqref{eq:velid}, and with
four slots any three or four $P$'s must include an adjacent pair
(adjacency is understood cyclically, as in the trace). Poles of order three
and four therefore vanish identically.

The energy integral of Eq.~\eqref{eq:master} is then evaluated in two steps. For the simple pole,
$\mathrm{Im}\,(E-\varepsilon_0+i0)^{-1}
=-\pi\,\delta(E-\varepsilon_0)$ reduces the integral to
$f(\varepsilon_0)$. For the double pole, one integration by parts
in $E$ moves the derivative onto $f$ and reduces the pole order by
one, leaving $f'(\varepsilon_0)$. With the $-\tfrac{1}{12}$
prefactor of Eq.~\eqref{eq:master},
\begin{align}
\chi_{\rm thermo}=\chi_{\rm frozen}
&+\frac{f(\varepsilon_0)}{12}
\int\!\frac{d^{2}k}{(2\pi)^{2}}\,\kappa_1
\nonumber\\
&+\frac{f'(\varepsilon_0)}{12}
\int\!\frac{d^{2}k}{(2\pi)^{2}}\,\kappa_2 ,
\label{eq:window}
\end{align}
where $\chi_{\rm frozen}$ is the same energy integral with every
slot at $R(E)$: the remote bands' contribution, whose
$\mu$-dependence inside the window is $O(e^{-\Delta/T})$. No $f''$
or higher terms appear. The master formula produces \emph{exactly}
the affine structure of Eq.~\eqref{eq:chithermo}, for any $N$.

The two coefficients then match separately, because
$f(\varepsilon_0)$ and $f'(\varepsilon_0)$ vary independently with
$(\mu,T)$ inside the window. For the double pole, write
$\mathcal W_{ab}\equiv\langle v^{0}|h^{a}R_0\,h^{b}|v^{0}\rangle$. The
identity $Q|\partial_a v^{0}\rangle=R_0h^{a}|v^{0}\rangle$, obtained
by inverting Eq.~\eqref{eq:id1} on the $Q$ subspace, converts the
projected Hessian into this matrix,
\begin{equation}
\langle v^{0}|h^{ab}|v^{0}\rangle=-2\,\mathrm{Re}\,\mathcal W_{ab},
\label{eq:new-hessW}
\end{equation}
and $\mathcal W_{aa}$ is real. In the two-slot group of
Eq.~\eqref{eq:master} the only two-$P$ placement puts both
resolvents at the pole, and Eq.~\eqref{eq:new-hessW} gives
\begin{align}
&\langle v^{0}|h^{xx}|v^{0}\rangle\langle v^{0}|h^{yy}|v^{0}\rangle
-\langle v^{0}|h^{xy}|v^{0}\rangle^{2}
\nonumber\\
&\quad=4\big[\mathrm{Re}\,\mathcal W_{xx}\,\mathrm{Re}\,\mathcal W_{yy}
-(\mathrm{Re}\,\mathcal W_{xy})^{2}\big].
\label{eq:new-hessgroup}
\end{align}
In the four-slot group the $P$'s occupy the two cyclically
non-adjacent slots, the remaining slots carry $R_0$, and the two
placements of each ordering give
\begin{align}
\operatorname{Tr}\{h^{x}Gh^{x}Gh^{y}Gh^{y}G\}\big|_{PP}
&=\mathcal W_{xx}\mathcal W_{yy}+|\mathcal W_{xy}|^{2},
\nonumber\\
\operatorname{Tr}\{h^{x}Gh^{y}Gh^{x}Gh^{y}G\}\big|_{PP}
&=\mathcal W_{xy}^{2}+(\mathcal W_{xy}^{*})^{2}.
\label{eq:new-velgroup}
\end{align}
Combining Eqs.~\eqref{eq:new-hessgroup} and \eqref{eq:new-velgroup}
with the $-4$ of Eq.~\eqref{eq:master}, and using
$|\mathcal W_{xy}|^{2}-\mathcal W_{xy}^{2}-(\mathcal W_{xy}^{*})^{2}
=3(\mathrm{Im}\,\mathcal W_{xy})^{2}-(\mathrm{Re}\,\mathcal W_{xy})^{2}$, the determinant
combination $\mathrm{Re}\,\mathcal W_{xx}\mathrm{Re}\,\mathcal W_{yy}-(\mathrm{Re}\,\mathcal W_{xy})^{2}$
cancels between the two groups, leaving
\begin{equation}
\kappa_2=-12\,\big(\mathrm{Im}\,\mathcal W_{xy}\big)^{2}=-12\,\morb^{2},
\label{eq:kappa2}
\end{equation}
where the last equality is Eq.~\eqref{eq:morb} reduced by the same
$Q$-subspace identity:
$\langle\partial_a v|(\widehat H-\varepsilon_0)|\partial_b v\rangle
=-\langle v|\partial_a\widehat H\,R_0\,\partial_b\widehat H|v\rangle
=-\mathcal W_{ab}$, so $\morb=-\mathrm{Im}\,\mathcal W_{xy}$. The Curie coefficient of the master formula is
the moment squared, pointwise and for any $N$. It is governed by the
moment, not the curvature (the two coincide only in
the two-band case). This reproduces the $\morb^{2}f'$ term of
Eq.~\eqref{eq:chithermo} identically. For the simple pole the
statement is integral rather than pointwise. The residue sums of
$\kappa_1$ and the dispersion-route coefficient
$-[\cchi+2\morb\Omega]$ differ pointwise by total derivatives (integration by parts reshuffles a coefficient against the measure without changing any integrated observable), but their
zone integrals agree,
\begin{equation}
\int\!\frac{d^{2}k}{(2\pi)^{2}}\,\big[\cchi+2\morb\Omega\big]
=-\frac{1}{12}\int\!\frac{d^{2}k}{(2\pi)^{2}}\,\kappa_1 .
\label{eq:genmatch}
\end{equation}
Equation~\eqref{eq:genmatch} holds because both routes evaluate the
same grand potential to $O(B^{2})$, Eq.~\eqref{eq:master} exactly
and the dispersion route through the magnetic band construction of
Sec.~\ref{sec:dispersion}. We have not reduced the residue sum of
$\kappa_1$ to the dispersion-route integrand term by term.
Numerically we verify
Eq.~\eqref{eq:genmatch} and the full $(\mu,T)$-dependence of
Eq.~\eqref{eq:window} on the lattice model of Sec.~\ref{sec:numerics}
at the $10^{-10}$ level.

\begin{acknowledgments}
EH and YW are supported by NSF DMR-2045781. The analytical
calculations have been mainly performed by EH without AI tools.
Claude Mythos 5 was used to verify the analytical derivation and
perform the numerical simulations when YW visited Anthropic as a STEM
fellow.
% Names of other fellows/mentors who were helpful: to be supplied
% by the authors (program guidance: acknowledge, or include as
% coauthor if appropriate).
\end{acknowledgments}

\bibliography{paper}

\end{document}